\documentclass[Journal,letterpaper, InsideFigs]{ascelike-new}
\usepackage[utf8]{inputenc}
\usepackage[T1]{fontenc}
\usepackage{lmodern}
\usepackage{graphicx}
\usepackage[figurename=Fig.,labelfont=bf,labelsep=period]{caption}
\usepackage{subcaption}
\usepackage{amsmath}
\usepackage{scalerel}
\usepackage{dirtytalk}
\usepackage{comment}
\usepackage{float}
\usepackage{titleref}

\usepackage{newtxtext,newtxmath}
\usepackage{setspace}
\usepackage[colorlinks=true,citecolor=red,linkcolor=black]{hyperref}
\NameTag{Simpson, \today}
\begin{document}
\nolinenumbers
\singlespacing
\title{A machine learning approach to Model Order Reduction of Nonlinear Systems via Autoencoder and LSTM Networks}

\author[1]{Thomas Simpson}
\author[2]{Nikolaos Dervilis, Ph.D.}
\author[1]{Eleni Chatzi, Ph.D.}

\affil[1]{Institute of Structural Engineering, Department of  Civil, Environmental and Geomatic Engineering, ETH Zürich, Stefano-Franscini Platz 5, 8093, Zürich, Switzerland. \\ Email: simpson@ibk.ethz.ch}
\affil[2]{Dynamics Research Group, Department of Mechanical Engineering, University of Sheffield. Sheffield S1 3JD, UK}

\maketitle

\begin{abstract}
In analyzing and assessing the condition of dynamical systems, it is necessary to account for nonlinearity. Recent advances in computation have rendered previously computationally infeasible analyses readily executable on common computer hardware. In certain use cases, however, such as uncertainty quantification or high precision real-time simulation, the computational cost remains a challenge. This necessitates adoption of reduced order modelling methods, which can reduce the computational toll of such nonlinear analyses. In this work, we propose a reduction scheme relying on the exploitation of an Autoencoder, as means to infer a latent space from output-only response data. This latent space, which in essence approximates the systems Nonlinear Normal Modes (NNMs), serves as an invertible reduction basis for the nonlinear system. The proposed machine learning framework is then complemented via use of Long-Short term Memory (LSTM) networks in the reduced space. These are used for creating a nonlinear Reduced Order Model (ROM) of the system, able to recreate the full system's dynamic response under a known driving input.
\end{abstract}

\section{Introduction}\label{introduction}
The use of nonlinear modelling in structural dynamics has become more common in recent years. Such methods provide higher precision when analyzing systems of higher complexity, as for example, those comprised of nonlinear materials or exhibiting large displacements. The practical use of such methods has largely been enabled owing to the increasing availability of powerful computational resources, which have made possible analyses that would have previously been prohibitively expensive. Certain use cases, however, remain wherein computational resources pose a limiting factor. Such problems are met in the domains of uncertainty quantification \cite{Sudret} or model updating \cite{Rogers_2019,EBRAHIMIAN2017194}, where repeated analyses of a system must be carried out, often with a real-time or near real-time requirement \cite{JOLDES20103305}. As such, the development of efficient reduced order modelling methods for nonlinear systems remains a necessity. In creating a Reduced Order Model (ROM) it is desirable to approximate a system to a certain degree of accuracy whilst reducing the cost of evaluation.

Reduction of finite element (FE) models came to prominence in the 1960s, mainly in the context of reduction of linear elastic systems via techniques such as the Craig-Bampton and MacNeal-Rubin methods \cite{Craig1968,DeKlerk2008}. Such methods are now widely used for the reduction of linear FE models \cite{Rixen2004,Papadimitriou} and respective implementations are incorporated in existing commercial analysis software \cite{Ansys,Abaqus}. These methods make use of linear normal modes as an optimal projection basis upon which the equations of motion can be reduced. The normal modes can easily be extracted either via an eigenvalue decomposition of the structural matrices, or statistically from output-only data \cite{SADHU2017415}. For linear systems these modes provide a very efficient reduction basis, as a result of the orthogonality of the normal modes. The use of these normal modes affords the additional possibility of targeting specific frequency ranges of interest, which is specifically useful in structural dynamics, where relevant system response is often constrained to the lower frequency region \cite{Besselink}.

When considering nonlinear systems, however, these linear methods are not appropriate, since the concept of linear normal modes no longer applies. This is a result of the presence of nonlinearity, which causes the modes to become energy dependent, thus eliminating the validity of superposition \cite{Kerschen2009}. The most common method for the reduction of weakly nonlinear FE models is Proper Orthogonal Decomposition (POD). In POD, time-series simulations of the system are carried out under appropriate forcing, with \say{snapshots} of the response field logged at various time intervals. These snapshots are then used to construct an approximate linear basis, for reduction of the nonlinear system through projection. This approximate basis is found through singular value decomposition of the response time-series snapshots \cite{Carlberg2009}. Related recent work explores the extension of the traditional reduction approach by inclusion of higher order enrichments, such as mode shape derivatives \cite{Wu2016}.

Beyond more conventionally used projection approaches, the further utilization of machine learning methods for constructing surrogates, or reduced order representations, has recently been an area of increasing interest for the dynamics community. Successful work has demonstrated the efficacy of neural network based methods, of varying architectures, for the metamodelling of nonlinear systems that comprise a relatively low number of degrees of freedom (DOFs) \cite{LAGAROS201292,Rih-Teng,ZHANG201955,Wu2011}. More recent work has explored the use of physics-informed machine learning methods and their benefits over purely data-based methods. Such physics-informed methods have been demonstrated in various contexts. Raissi et al. \cite{Raissi} employ deep neural networks to tackle a broad class of nonlinear partial differential equations. These deep neural networks are differentiated using automatic differentiation and are further exploited to control the dynamics of the system. A similar method has been demonstrated for dealing with axial compression buckling analysis of a thin walled cylinder \cite{Tao}. In the area of structural dynamics, Zhang et al. \cite{ZHANG2020113226} represent a nonlinear structural system as a state-space model whose transfer function and the calculation of the restoring force vector are represented using Long Short Term Memory (LSTM) neural networks, showing improvements over an alternative non physics-informed metamodel. In the domain of nonlinear aerodynamics, Snaiki and Wu \cite{Snaiki2019} make use of physical equations for the regularisation of LSTM neural networks used to metamodel nonlinear aerodynamic simulations. An interesting work, making use of  networks for the reduced order modelling of structural systems with nonlinear aerodynamic effects, has also been demonstrated by Wang and Wu \cite{Wang2020} wherein physical knowledge of the system has been used to aid in regularisation of the networks. A more thorough description of LSTM networks and a review of their usage in dynamics is given in Section: Recurrent Neural Networks. Once again, however, these works have largely been limited to a low number of DOFs. A problem here is that an increase in the number of DOFs tends to result in a corollary increase in the required number of neural network parameters. This can cause difficulties with regards to training and overfitting.

Nonlinear Normal Modes (NNMs) extend the concept of normal modes to nonlinear systems and can provide an alternative approach for reduction bases. Several formulations of NNMs exist, with the most common being the Shaw-Pierre formulation \cite{Shaw1993}, which builds on the fundamental work of Rosenberg \cite{Rosenberg1962}. These NNMs in essence represent a nonlinear curved manifold, which delivers a more efficient reduction basis for nonlinear systems \cite{Haller2016}. The capturing of manifold curvature results in these NNMs being valid over the whole phase-space of a system. A linear scheme, on the other hand, is limited in that the energy dependence of the mode shapes requires multiple reduction bases for different energy levels. However, considerable difficulty  remains in terms of the extraction of these NNMs. Both analytical and numerical methods have been formulated for the purpose of determining the NNMs of given systems. However, the majority of these methods remain limited to reasonably simple systems, while the employed formulations tend to depend on the type (characteristics) of the underlying nonlinearity. The premise of this work is that extraction of NNMs via statistical methods on the basis of output-only data, as demonstrated in recent works \cite{Worden2016,Dervilis2019}, could facilitate extraction of NNMs from a broader range of systems and with little analytical work required.

The extraction of NNMs by means of a nonlinear transform raises, however, a further problem. Whilst in projection methods the equations of motion are projected onto the reduced coordinate set and are then solved through integration, it is unclear how these equations could be used after the application of an arbitrary nonlinear transform. We propose the use of a statistical regression model, which can learn the dynamics of the system within the reduced nonlinearly transformed space. This requires the training of the regression with sufficient input and output data for representative forcing types.

In this work, a reduced order modelling approach is demonstrated, aiming to capitalize on approximation of NNMs as a reduction basis. A nonlinear reduced space, yielding NNM-like quantities, is extracted from output-only data using autoencoder neural networks. A recurrent neural network (RNN) is then used to construct a dynamic regression model on the derived NNM basis. This trained regression model is coupled with the decoder functionality in order to reproduce the response the full order system. This method is demonstrated on several different nonlinear systems of varying complexity and nonlinearity types.

The layout of this paper is structured as follows: Section: Nonlinear Normal Modes lays down the relevant background on NNMs, their extraction and use in ROMs. Section: Machine Learning Framework presents the machine learning methods used for construction of the ROM. Section: Autoencoders describes autoencoder neural networks and how these are used herein for extraction of NNM-like quantities. Section: Recurrent Neural Networks presents an overview of recurrent neural networks (RNN) and further elaborates on the LSTM networks used herein for learning the system dynamics. Section: ROM Using Autoencoders offers the further details of the developed ROM framework, including the different stages of training and testing of the ROM. Section: Examples presents 3 exemplar nonlinear systems and the respective performance of the ROM method. Section: Physical Interpretation of NNMs discusses some interesting features of the NNMs as extracted by the autoencoder networks. Finally, Section: Conclusions and Further Work presents the conclusions drawn from this work and suggested future developments.

\newpage
\section{Nonlinear Normal Modes}\label{sec:NNMs}
\subsection{Shaw-Pierre Manifolds}
Shaw-Pierre NNMs were first proposed in 1993. They defined an NNM as \say{a motion which takes place on a two dimensional invariant manifold, formulated in the phase-space}. This implies that the motion of all of the individual DOFs in the system is described as a fixed function of the displacement and velocity of a single DOF. This function, if plotted in the phase-space, would form a surface which defines the mentioned manifold. In the case of linear systems, these manifolds are simply planes, which describe the constant amplitude ratio between points in a linear mode. Nonlinearities in a system cause curvature of this plane to form a manifold. This curvature is a result of the energy dependence of mode shapes  observed in nonlinear systems \cite{Shaw1993}.

Calculation of NNMs can, however, be difficult. The assumption of an NNM being an invariant manifold is represented by equation \ref{eq:SP1}. In this equation, $x_i$ and $\dot{x_i}$ denote the displacement and velocity of the i-th coordinate of the response, whilst $u_i$ and $v_i$ represent the displacement and velocity of the i-th transformed coordinate in the NNM. The velocity and displacement of each of the coordinates in the NNMs are described as a function of a displacement-velocity pair from a single master coordinate. These functions are denoted as $f_i$ and $g_i$. This relationship is then substituted into the equation of motion, which results in a new formulation. This new formulation is, however, as difficult to solve as the original case. The solution here is to assume a power series expansion for the functions $f_1\cdots f_n$ and $g_1\cdots g_n$. This then allows for an analytical solution in some cases and a numerical solution for the general case.

\singlespacing
\begin{equation}\label{eq:SP1}
\begin{bmatrix}
x_1\\\dot{x}_1\\x_2\\\dot{x}_2\\\vdots\\x_n\\\dot{x}_n
\end{bmatrix}=
\begin{bmatrix}
u_1\\v_1\\u_2\\v_2\\\vdots\\u_n\\v_n
\end{bmatrix}=
\begin{bmatrix}
u_1\\v_1\\f_2(u_1,v_1)\\g_2(u_1,v_1)\\\vdots\\f_n(u_1,v_1)\\g_n(u_1,v_1)
\end{bmatrix}
\end{equation}

\subsection{Output-Only Extraction}

Previous studies have also examined use of machine learning methods for output-only extraction of NNMs from nonlinear systems. The driving idea being that, similarly to POD, time-series response data is extracted and a reduction basis is formed via extraction of a latent space. Worden and Green \cite{Worden2016} represented NNMs as a truncated series of polynomials, similarly to the Shaw-Pierre formulation, with the coefficients of these polynomials being identified via the use of an optimisation procedure. During optimisation, statistical independence of the NNMs is also enforced, up to a given order. The enforcing of independence serves to ensure the modal invariance property, which should be inherent to NNMs. Modal invariance implies that motion within one NNM should not influence motion within another NNM. In the work of Dervilis et al. \cite{Dervilis2019}, the extraction of NNMs from output-only data using a category of machine learning algorithms known as manifold learning methods is demonstrated. Manifold learning encapsulates a class of methods used to extract a lower dimensional, possibly nonlinear structure from high dimensional data. This work focused on the locally linear embedding algorithm (LLE). The method was demonstrated initially on simulated datasets of systems, exhibiting cubic nonlinearities. It was further demonstrated on a 4 DOF experimental dataset with a saturation type nonlinearity, produced by the Los Alamos national laboratory \cite{Figueiredo2009}. These works were, however, fairly introductory and have focused mostly on finding the transform from the physical space to the NNMs and the inverse operation. The utility of these extracted transforms for reduced order modelling and the extension to systems of a greater number of DOFs is yet to be explored. 

\subsection{Usage in ROMs}
The use of NNMs for reduced order modelling has been broadly exploited in the past few decades, with the majority of published works focusing on relatively small scale examples and the physical interpretability of the NNM reduction, mostly in the sense of nonlinear modal analysis. For example, numerous methods focus on the extraction of backbone curves for nonlinear systems. The backbone curves can be used to characterise and quantify nonlinearities in the system and to inform reduced order models \cite{LiuWagg}. A detailed review of the theoretical basis and validity of NNMs for reduced order modelling is given in \cite{Haller2016}. Various work has shown the effectiveness of NNMs as a basis for reduced order models, as reported in \cite{TOUZE2006958,PESHECK20011085,Kuether2015,JAIN2018195}. In these works a mixture of analytical and numerical methods are adopted in the formation of the ROMs. In all demonstrated examples, a considerable amount of analytical analysis or manipulation of the FE models is required. Furthermore, for the most part, these works are relatively limited in terms of the dimensionality of the models considered. Moving away from this restriction, it is notable that in \cite{Kuether2015} a large FE model is considered, which is reduced to successfully recreate the spectral behaviour of the systems NNMs. This analysis is, however, limited to spectral properties and does not consider time-series response reconstruction or prediction.

\section{Machine Learning Framework}\label{sec:machinelearning}

\subsection{Autoencoders}\label{sec:autoencoders}

As previously mentioned, it is thought that manifold learning methods may be exploited for output-only extraction of NNMs. Autoencoder (AE) neural networks are one such manifold learning technique, often used for dimensionality reduction or de-noising problems \cite{Hinton2006}. They are constructed, as illustrated in Figure 1, as a deep neural network architecture with a so-called \say{bottleneck} layer, $X_i$ represents the input vectors to the network, $Z_i$ the latent space vectors and $\Tilde{X}_i$ are the output vectors which approximate the input vectors. This bottleneck layer generally has a smaller number of nodes than the input and output layers of the network, with, in the case of the autoencoder depicted in Figure 1, 2 nodes in the bottleneck layer. The key concept of an autoencoder lies in the utilisation of a cost function, which attempts to recreate the inputs as closely as possible at the output. The bottleneck layer forces the data through a lower dimensional representation in its attempt to recreate the inputs. This gives a near-optimal reduction of the data onto the chosen number of dimensions. The hidden layers in the network allow for a nonlinear transform of the inputs before the bottleneck layer. In training an AE, both the encoding operation, which transforms the physical to the latent space, is learned as well as the decoding operation, which returns to the physical space from the latent space.

\begin{figure}[h!]
    \centering
    \includegraphics[width=110mm]{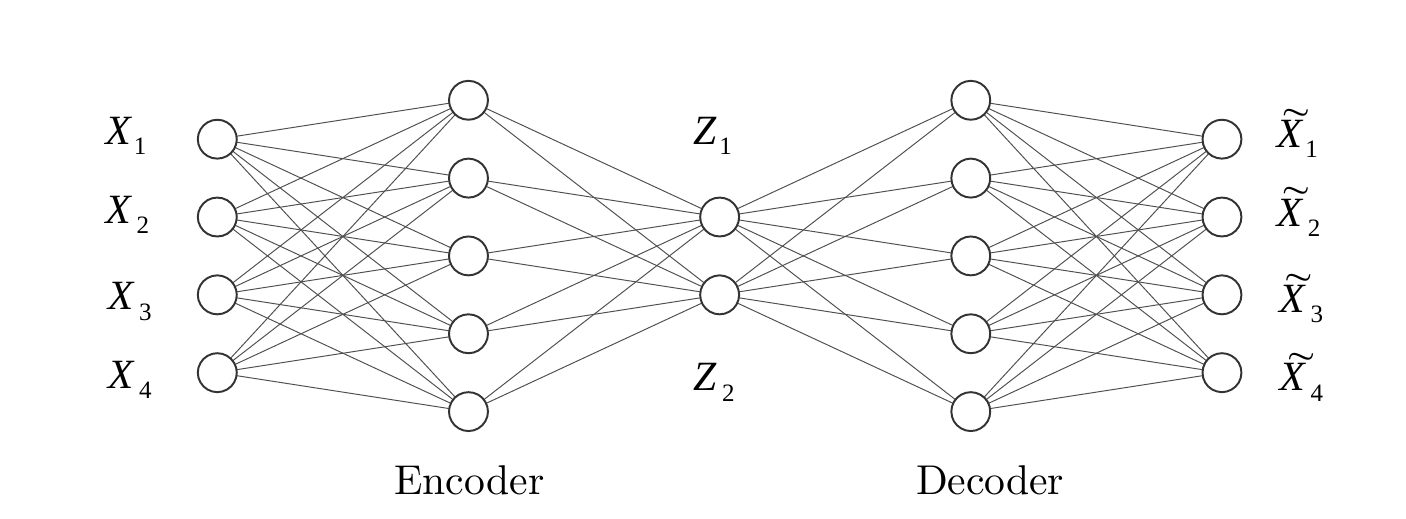}
    \caption{Architecture of an AE with 4 dimensional input/output and a 2 dimensional bottleneck.}
    \label{fig:AE}
\end{figure}

Autoencoders are often used in a machine learning context as unsupervised feature extractors, for extracting more robust lower dimensional features from high dimensional data \cite{Vincent2008}. They form part of a group of algorithms known as manifold learning algorithms. These are algorithms which attempt to obtain structure from data; perhaps the most popular such algorithm being the linear method principal component analysis (PCA). PCA is a method which finds the optimal orthogonal projection of a dataset onto a reduced number of dimensions, whilst preserving the maximum possible variance \cite{WOLD198737}. Nonlinear manifold learning methods essentially aim to perform the same task as PCA by determining an optimal nonlinear embedding of the data on a reduced number of coordinates. When considering nonlinear manifold learning, autoencoders offer numerous compelling advantages. Firstly, the use of a neural network architecture allows for scalability to very large dimensions, which is more difficult with certain other methods, such as locally linear embedding \cite{Roweis2323}. Secondly, unlike other methods, autoencoders naturally provide the reverse transform, which can return from the reduced coordinate space to the global coordinates. This is an essential property for ROM in structural dynamics as it effectively allows the physical response of the structure to be recovered from the NNM response. Autoencoders have been used in a number of dynamic systems where reduction on the basis of a manifold of lower dimension is deemed feasible \cite{Yoo2017}. In Holden et al. \cite{motionManifold} autoencoders are also used to capture a lower dimensional representation of videos of human motion. In a dynamics context, Lee and Carlberg \cite{LEE2020108973} made use of convolutional autoencoders in conjunction with the nonlinear Galerkin method to construct reduced order models. The convolutional autoencoder was used to characterise the underlying manifold of the solution and hence provide local projection basis for the ROM.

In the case of nonlinear structural dynamics,  the lower dimensional manifold extracted by the autoencoder could correspond to the NNMs of the system \cite{Dervilis2019,Worden2016}. Similarly to how a greatly reduced number of linear mode shapes can often be used to reconstruct the approximate response of a high dimensional linear FE model. The commonly used POD method has been shown to extract the best linear approximation of NNMs from data \cite{Kerschen2005}. An autoencoder with linear activation functions has also been shown to be equivalent to POD \cite{BALDI198953}. Kerschen et al. remark that the use of a nonlinear autoencoder gives an interesting nonlinear extension to POD. As, in this work, we do not require a strict definition of NNMs for the construction of the ROM, we consider the autoencoder to reflect a better approximation of NNM-like quantities than the POD approximation. The encoder segment of an autoencoder is the function through which the high dimensional input data can be transformed to the lower dimensional manifold or NNM space, whilst the decoder segment provides the inverse operation. This inverse operation of transforming from the latent back to the physical space can be seen as equivalent to the nonlinear superposition of NNMs demonstrated in \cite{Shaw1993}. Due to the nonlinear activation functions, the encoding and decoding operations can both be nonlinear functions and hence allow for the curved nature of the manifolds to be represented.

\subsection{Recurrent Neural Networks}\label{sec:RNN}
As previously mentioned, the use of a nonlinear transform to extract NNMs means that we can no longer make use of the equations of motion in a simple manner. As such, a statistical regression must be used to represent the dynamics of the system on the NNM space. Recurrent Neural Networks (RNNs) are a class of architectures developed specifically for dealing with sequence data, such as time-series \cite{Rumelhart}. A basic \say{vanilla} RNN uses a unit, which admits as input a concatenation of the exogenous input and of the hidden states of the network, which is propagated from the previous time step . From this input, a new value of the hidden state layer is calculated through a traditional feed-forward neural network layer. This is carried out through multiplication of the input vector with a weights matrix, the addition of a bias term and the application of an activation function. It is important to note that in an RNN the weight matrix and bias terms used for calculating the hidden state are shared between all time steps of the RNN. The output at each time step can then be calculated through an additional layer, or layers, which act upon the hidden state of the RNN. The passing of hidden state information from the previous time step is what allows the RNN to capture time dependencies. This can be viewed as being similar to a nonlinear autoregressive with exogenous input (NARX) model; more traditionally used in system identification \cite{NOEL20172}, in which the regressor functions are replaced by a neural network. A key difference to a NARX model, however, is that rather than explicitly passing the output at the previous time as input, RNNs have the freedom to pass additional information in the form of the hidden state. Further advantages of using RNN architectures include the easy and natural handling of multiple input multiple output (MIMO) regression. Furthermore, the extensive use of RNNs in machine learning research has resulted in very efficient implementations and learning algorithms, which help when scaling to large dimensional problems and large amounts of data. The basic architecture of an RNN is demonstrated in Figure 2. In this figure, $X_i$ represents the exogenous input vector at time $i$, $Y_i$ the network output at time $i$, and $A_i$ the hidden state vector of the network at time $i$ which is passed as an additional input to the network at time step $i+1$.

\begin{figure}[h!]
    \centering
    \centerline{\includegraphics[width=150mm,trim={0cm 0 0cm 0},clip]{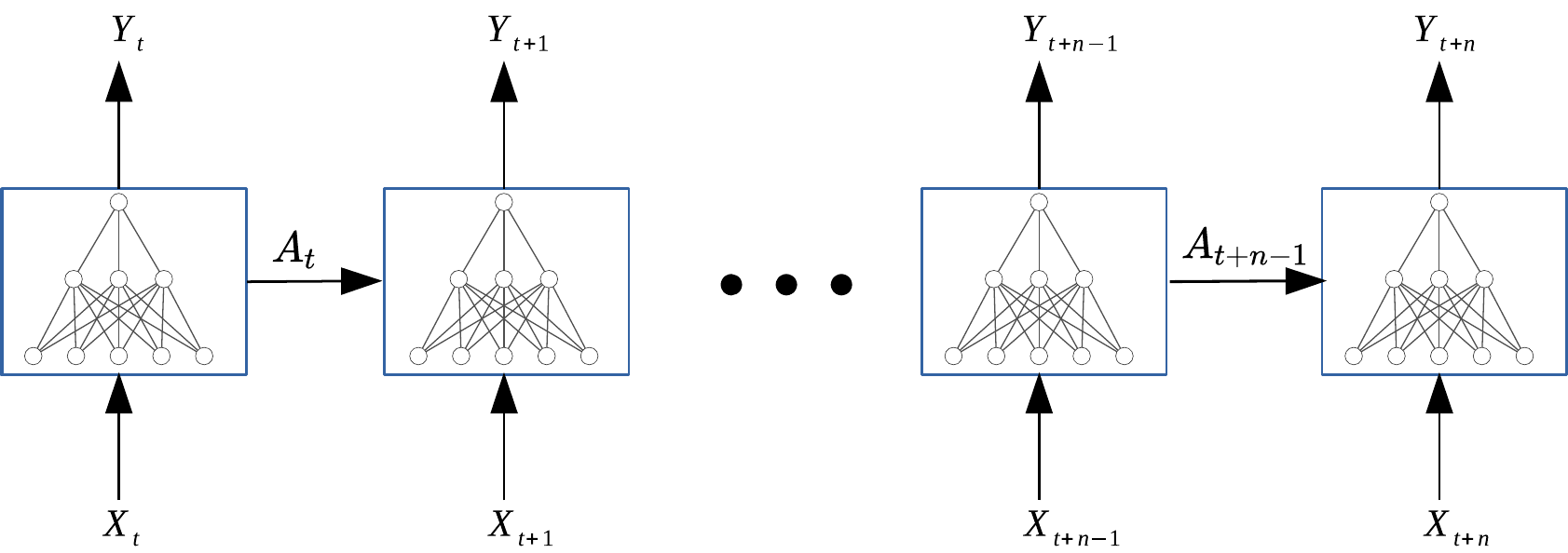}}
    \caption{Architecture of a recurrent neural network with the state connections passing information through time.}
    \label{fig:RNN}
\end{figure}
 
 Theoretically, RNNs can be very powerful at learning relationships between time-series with long term dependencies. However, in practice it is found that when training an RNN using the back propagation through time algorithm, that the process suffers from vanishing or exploding gradient problems. This means that as the error gradients are propagated back through many time steps, the gradients exponentially decrease to zero or increase towards infinity. As a result of this, a conventional RNN struggles to propagate error gradients back through multiple time steps and hence fails to learn long term dependencies \cite{Bengio1994}. This motivated the development of more advanced RNN cells, such as the long short term memory (LSTM) cell \cite{Hochreiter1997}.

\subsubsection{Long Short Term Memory Networks}
Long short term memory (LSTM) networks are a more powerful variant of an RNN, developed to overcome the vanishing/exploding gradient problem \cite{Hochreiter1997} and are responsible for much of the success of RNNs in recent years. The key innovation of LSTM networks is the use of gated activation functions, which control the updating of the \say{Cell state}, which is analogous to the network memory. This allows LSTM networks to better capture long-term dependencies in sequences and has lead to significant success in disparate fields, such as sequence modeling \cite{Bayer2015phd}, handwriting recognition \cite{Graves2007} and language translation \cite{Wuetal2016}. There exists, however, only a limited amount of work that exploits LSTM networks in a dynamical system identification context. Vlachas et al. \cite{vlachas2018a} consider the use of an LSTM network to model a chaotic Lorenz dynamic system, whilst Wang \cite{Wang2017} demonstrates the use in a control systems context to identify a plant model. The use of LSTM networks has also been demonstrated for the reduced order modelling of nonlinear aerodynamic simulations \cite{Li2020,Wang2020}. Very recent work has also made use of physics informed LSTM networks to improve performance compared to purely data driven methods \cite{ZHANG2020113226}.

Much of the success of LSTM networks has been ascribed to their ability to \say{remember} information for arbitrary periods of time. One, oft cited, example of this ability is found in the context of natural language processing and pertains to the ability to store contextual information, such as the tense or gender of a sentence subject, which is required to correctly conjugate a verb when significantly separated within a sentence. This arbitrary memory could be of considerable interest in structural dynamics problems. Many nonlinear models considered in nonlinear dynamics exhibit such a long term memory behaviour. For example, the hidden state variable in the Bouc-Wen hysteresis model \cite{Ismail2009TheHB}.

The key to the LSTM network is the Cell state. This cell state acts as the memory of the network and allows the passing of information through long sequences. An LSTM network uses 3 \say{gates} to control this cell state, the forget gate, the update gate and the output gate. It is as a result of these gate functions that the LSTM network proves resistant to vanishing and exploding gradients. Due to the gating functions, the gradient of the cell is no longer intrinsically driven towards zero as in standard RNNs preventing vanishing gradients and the gradient of the cell state is upper bounded by 1 preventing exploding gradients \cite{Bayer2015phd}. This gives the LSTM network architecture much better capability to learn long term relationships than a vanilla RNN cell.
The architecture of the LSTM cell along with the governing equations are shown in Figure 3. $i_t$, $f_t$, $o_t$ are the gates signals, of the forget, input and, output gates respectively. $X_t$, $h_{t-1}$, $C_{t-1}$ are the exogenous input, and the hidden and cell state from the previous time step whilst $U^i$, $U^f$, $U^0$, $W^i$, $W^f$, $W^0$ are weight matrices. The activation function $\sigma$ is a sigmoid function. The full description and function of these equations can be found in the literature \cite{Hochreiter1997}.

\begin{figure}[h!]
    \centering
    \includegraphics[trim={0mm 0mm 0mm 0mm},clip,width=150mm]{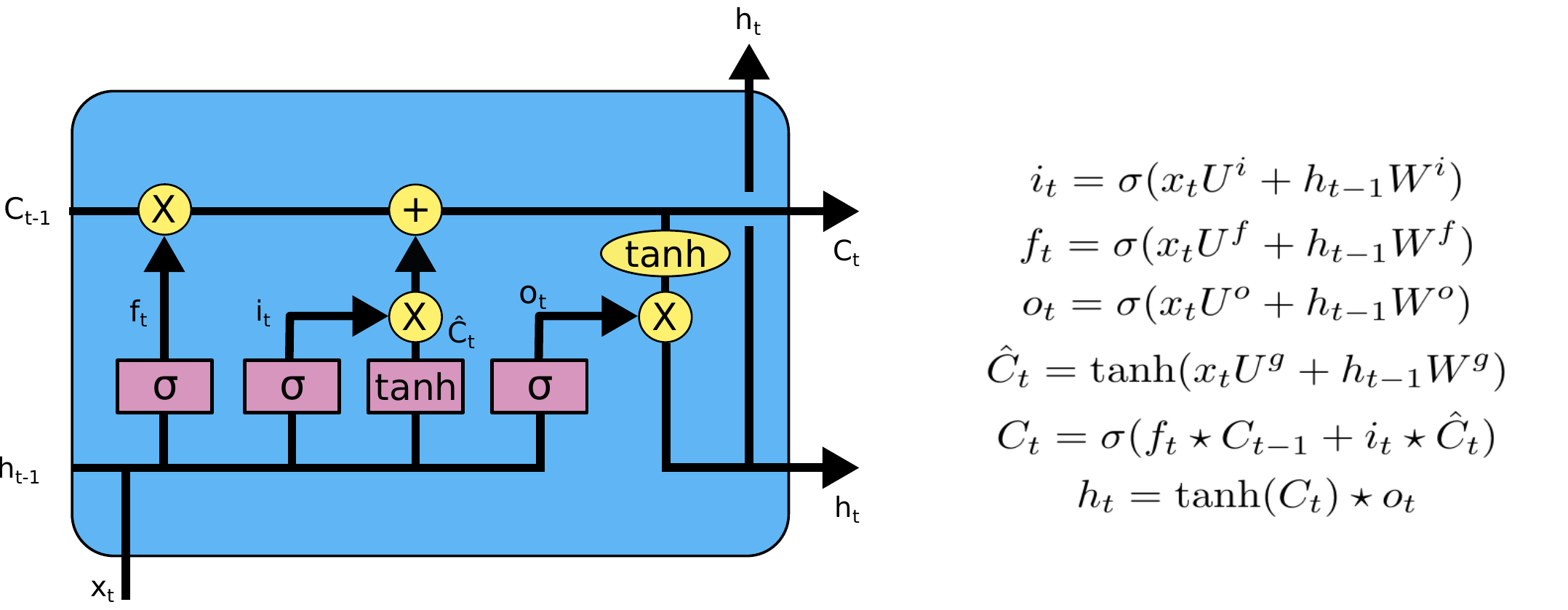}
    \caption{Architecture of the LSTM network cell, adapted from \protect\cite{COLAH}}
    \label{fig:LSTM}
\end{figure}

\section{ROM using Autoencoders}\label{sec:ROM}
In this section the reduced order modelling method developed herein is described in detail. The first step is to generate training data from the system of interest. This is achieved by performing time-series simulations of the system under forcing that is representative of the forcing of interest. Both the input and output time-series of these simulations are then stored for all system DOFs. The autoencoder is trained for extracting NNMs using only the output response data of the system. The autoencoder compresses the dimensionality of the data to a reduced number of latent variables, with this reduced number being the number of NNMs retained for the ROM. This number is selected by balancing the amount of reduction against the fidelity of the ROM. This training procedure is described in Figure 4, where vectors $X_i$ represent the time histories of response from each of the DOFs in the system, which are collected in matrix $X$. In this process the cost function used is described by Equation \ref{eq:MSE}, where $M$ denotes the number of DOFs in the system, $N$ stands for the number of data points or time steps, $X_i$ is the $M$ dimensional true response vector at step $i$ and $\hat{X_i}$ is the approximated $M$ dimensional response vector outputted by the autoencoder for the sample at step $i$. This corresponds to the mean squared error of reconstruction from the autoencoder with reduced number of NNMs.

\begin{equation}\label{eq:MSE}
    \ell(\hat{x})=\frac{1}{MN}\sum_{j=1}^{M}\sum_{i=1}^{N}(X_i-\hat{X_i})^2
\end{equation}

\begin{figure}[h!]
    \centering
    \centerline{\includegraphics[trim={0mm 0mm 0mm 0mm},clip,width=165mm]{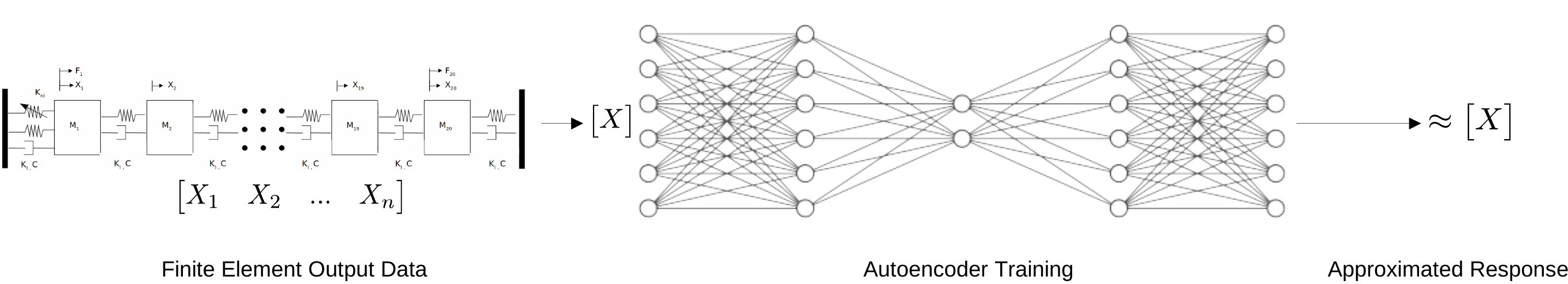}}
    \caption{1st step of the proposed ROM framework; a reduced number of NNMs are extracted from output-only dataset.}
    \label{fig:ROM1}
\end{figure}

\begin{figure}[h!]
    \centering
    \includegraphics[width=165mm]{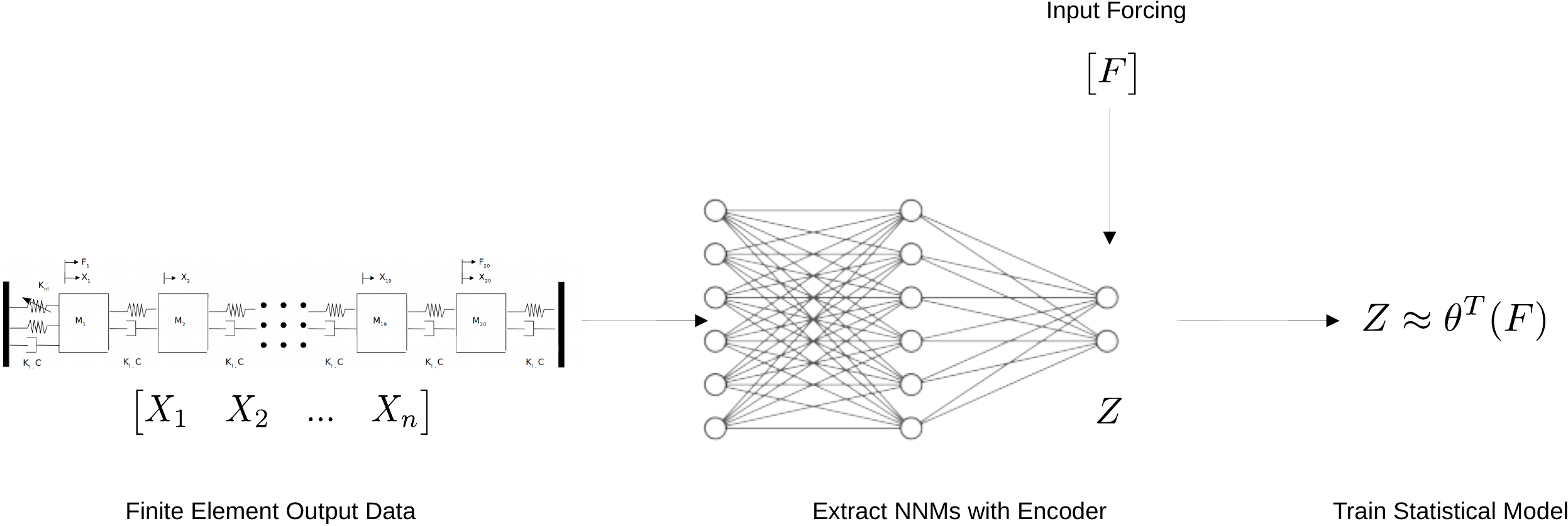}
    \caption{2nd step of the proposed ROM framework; a regression model is trained which predicts response in the retained NNMs based on forcing histories.}
    \label{fig:ROM2}
\end{figure}
\begin{figure}[h!]
    \centering
     \includegraphics[width=165mm]{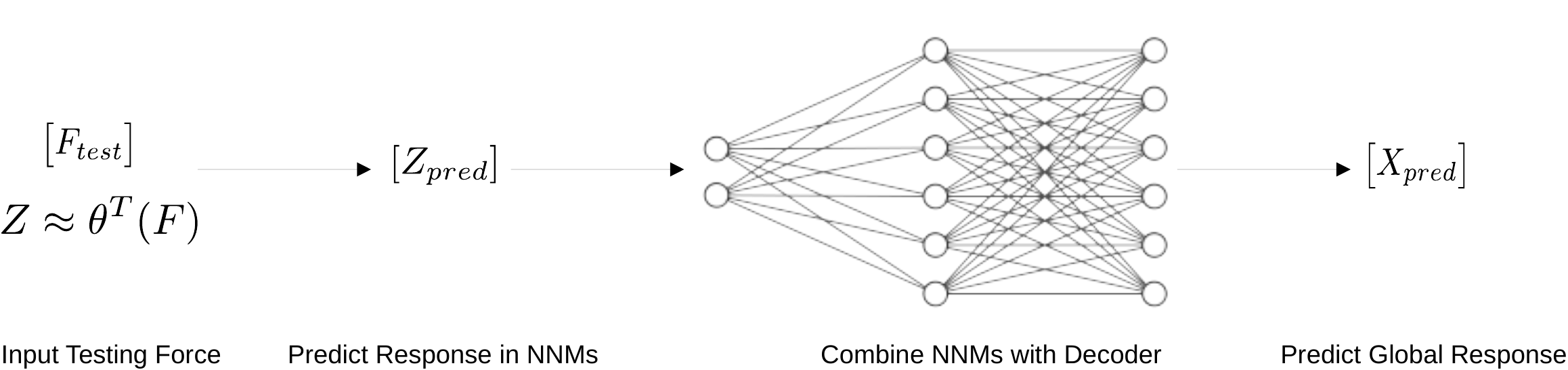}
    \caption{3rd step of the proposed ROM framework; the response in the NNMs is predicted for new forcing histories and the full response is reconstructed through the decoder.}
    \label{fig:ROM3}
\end{figure}

In contrast to projection-based reduction methods, wherein the original equations of motion can be retained and solved on the reduced coordinate space, the use of the nonlinear transform herein requires a different approach. In this case, a statistical regression method is used to learn the dynamics of the system response within the NNM space. A statistical regression is trained, which can predict the response of the system in the latent space based on new forcing time histories, as illustrated in Figure 5. The regression model, $g(F)$, is trained based on the forcing time histories used for the simulation, $F$, and the corresponding response in the NNMs as extracted by the trained encoder, $Z$.

The final step in the metamodelling process lies in predicting the system's response for a new applied forcing time history. As illustrated in Figure 6, the trained regression model, $g(F_{test})$ is used to predict the response of the system within the NNMs, to the new forcing time histories $F_{test}$. Having predicted this response, $Y_{pred}$, the full field response, $X_{pred}$, can be recovered by using the decoding portion of the autoencoder.

\section{Examples}\label{sec:Examples}
To demonstrate the utility of the method, various nonlinear systems were considered in order to investigate whether they could successfully be emulated by such a modelling scheme. First a relatively simple, 20 DOF system was considered featuring cubic type nonlinearity. This 20 DOF system was considered both with only a single source of nonlinearity and subsequently with multiple sources. Secondary to this, a larger, 108 DOF system, representing a building frame, was considered. The nonlinearity in this second system assumes the form of a Bouc-Wen modelled hysteresis.

\subsection{Single Nonlinearity}

We verify the applicability of the proposed method on a 20 DOF mass, spring, damper system. The system was considered to be comprised of 20 masses in a chain, with linear spring and linear damping elements between each mass, and additionally a cubic spring between the first mass and the boundary, as demonstrated in Figure 7. The nonlinearity of the nonlinear spring elements is of the cubic type. Forcing was applied to this system in two locations, at the first and final masses in the chain. All the parameters are assumed uniform across all DOFs and the underlying linear system is proportionally damped; the parameters are defined as follows:  $M_i = 0.1\,kg$, $K_l = 100\,N/m$, $C_l = 0.1\,kg/s$, $K_{nl} = 2500\,N/m^{3}$. 

\begin{figure}[h!]
    \centering
    \includegraphics[width=150mm]{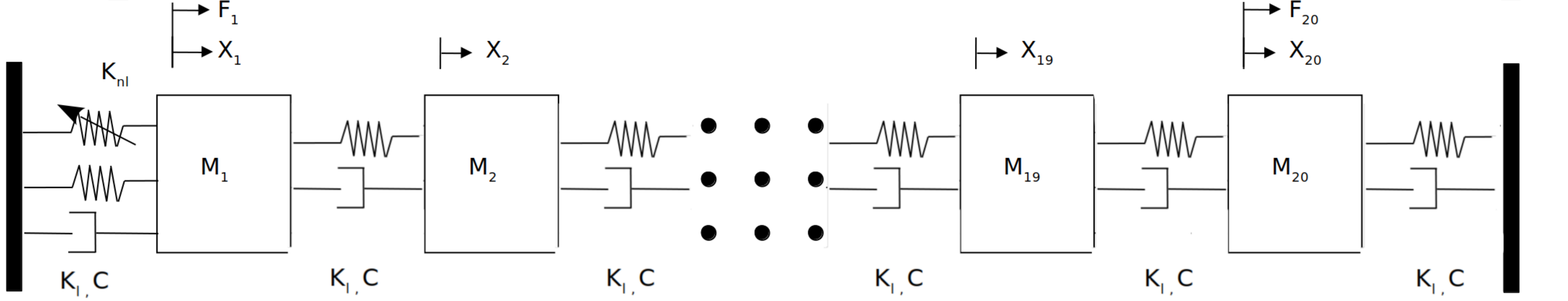}
    \caption{20 Degree of freedom mass spring damper system with a single nonlinearity between the ground and the first mass.}
    \label{fig:20dofoneNonLin}
\end{figure}

Forcing was applied to the system at each of the end DOFs as a low-pass filtered white noise signal with different realisations of the signal applied at either end. Each forcing signal corresponds to a variance of $24 N^2$. The cut-off frequency of the low pass filter was set at 7.5 Hz, which lies between the 11th and 12th natural frequencies of the linearised system. This cut-off frequency was selected so as to focus the excitation on the structurally important lower frequency modes. The time-series solution of the system was performed using the 4th order Runge-Kutta integration scheme at a time step of $0.01 s$. Figure 8 compares the magnitude of the restoring force in the linear spring and the nonlinear spring connecting the first mass to ground. The magnitude of the nonlinear restoring force is comparable and at times even larger than the linear spring indicating that this is indeed a significantly nonlinear system. This is also indicated by Figure 9 which plots the total restoring force against displacement at the first degree of freedom. The restoring force demonstrates a clear strain hardening nonlinear behaviour.

\begin{figure}[h!]
    \centering
    \includegraphics[width=150mm]{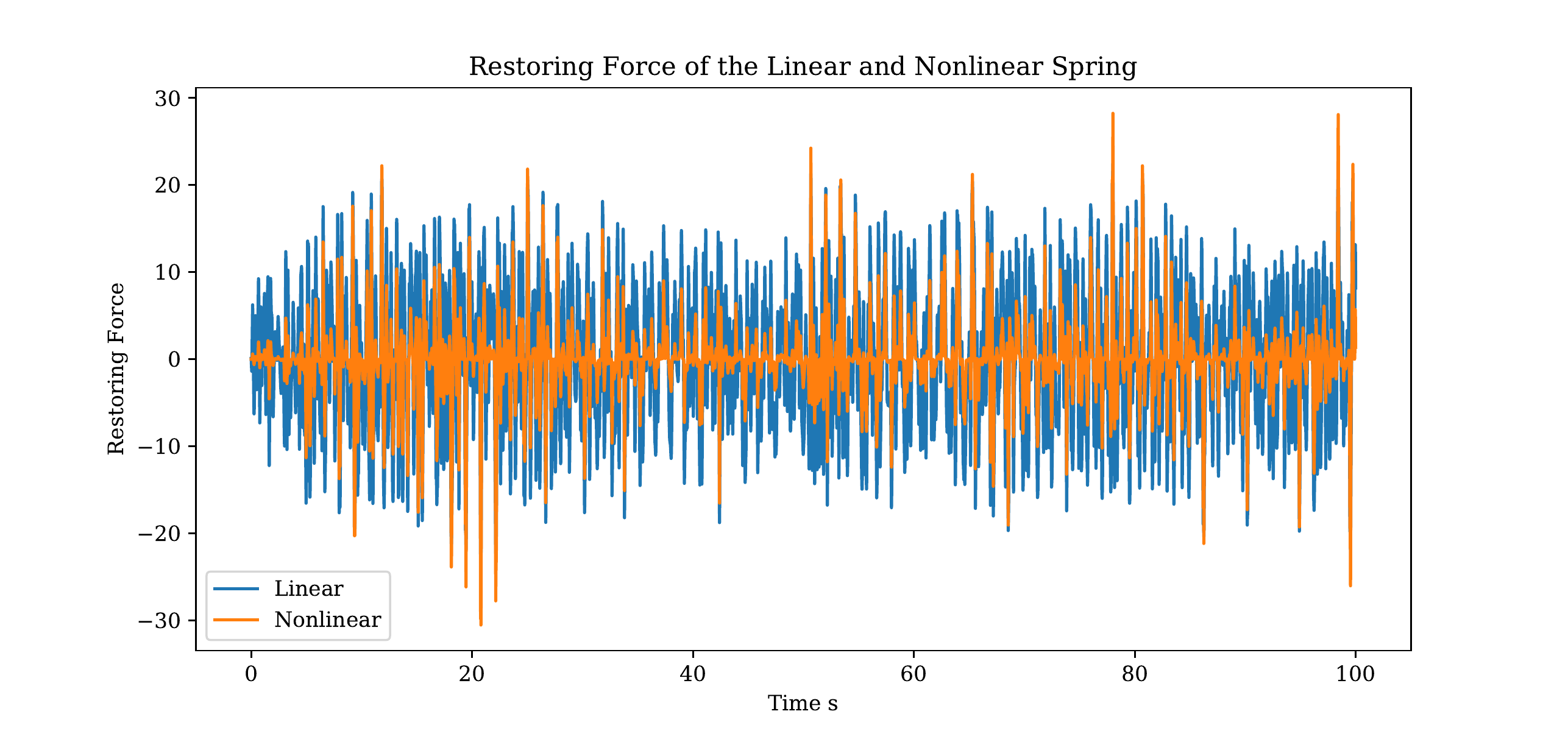}
    \caption{Relative magnitude of the linear and nonlinear restoring forces at the first degree of freedom.}
    \label{fig:RestForce1NL}
\end{figure}

\begin{figure}[h!]
    \centering
    \includegraphics[width=90mm]{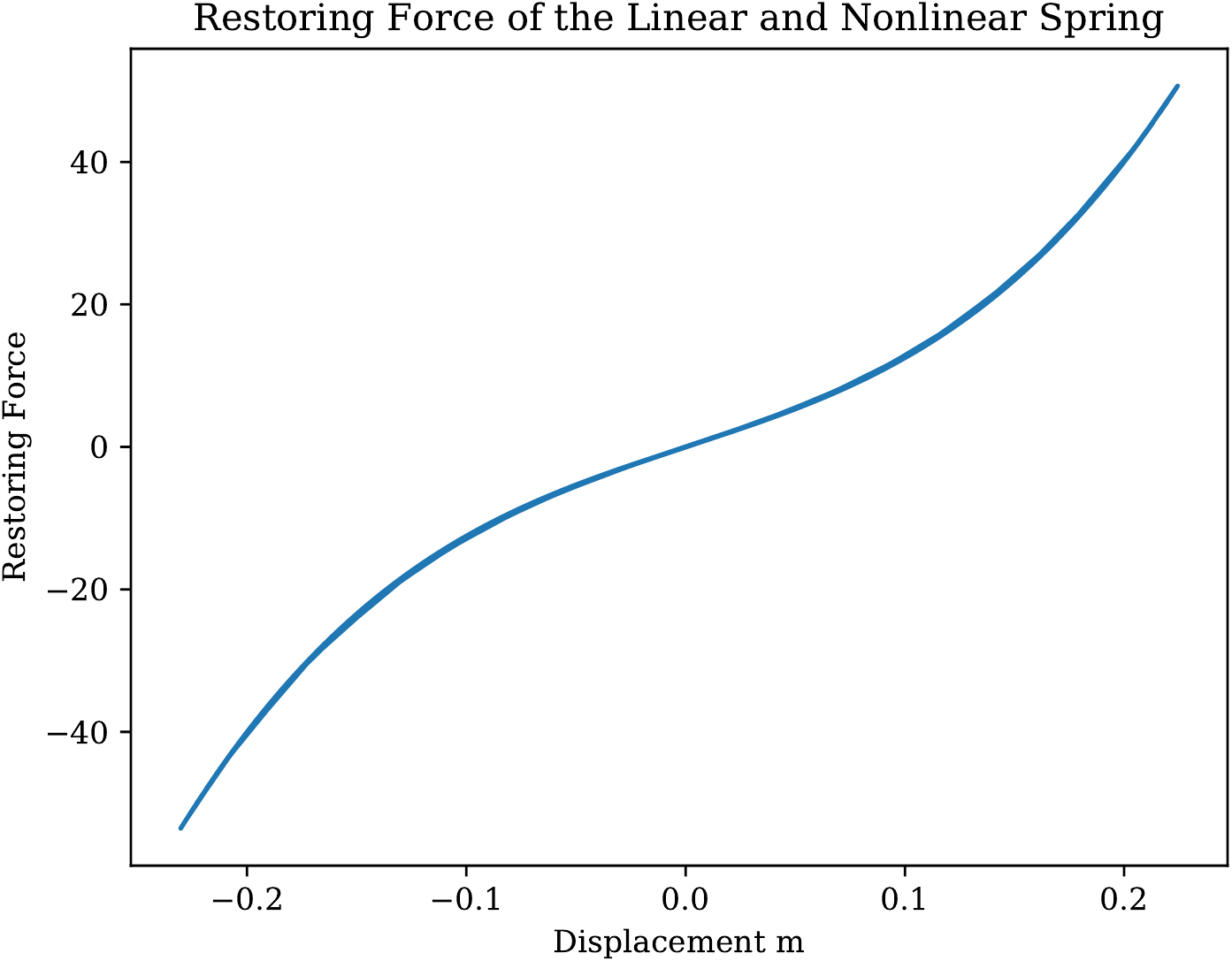}
    \caption{Force-Displacement plot of the restoring force at the first degree of freedom.}
    \label{fig:DispForceOne}
\end{figure}

\clearpage
\subsubsection{Reconstruction}
For the first step of constructing the ROM, the displacement time histories were extracted from simulations and an autoencoder was then trained with a reduced number of hidden units in the bottleneck layer. This number of hidden units corresponds to the number of retained NNMs and hence defines the amount of reduction of the system. Various autoencoders were trained with the number of hidden units varied from 1 to 20. Figure 10 shows the mean squared error of reconstruction achieved by each of the autoencoders for varying number of hidden units. Such a graph can aid in picking an appropriate number of NNMs to be retained although this method may not be suitable for very high dimensional data, where the cost associated with autoencoder training could be prohibitive. The mean squared error of reconstruction reduces with increased number of NNMs, as expected. However, in some cases, retaining an increased number of NNMs results in a reduction in the fidelity of reconstruction. This is a result of the training process of neural networks, since the gradient descent algorithm is not always guaranteed to determine a global minimum. As such, in some cases a more complex network, which ought to yield improved performance, shows inferior performance.

\begin{figure}[h!]
    \centering
    \includegraphics[width=100mm]{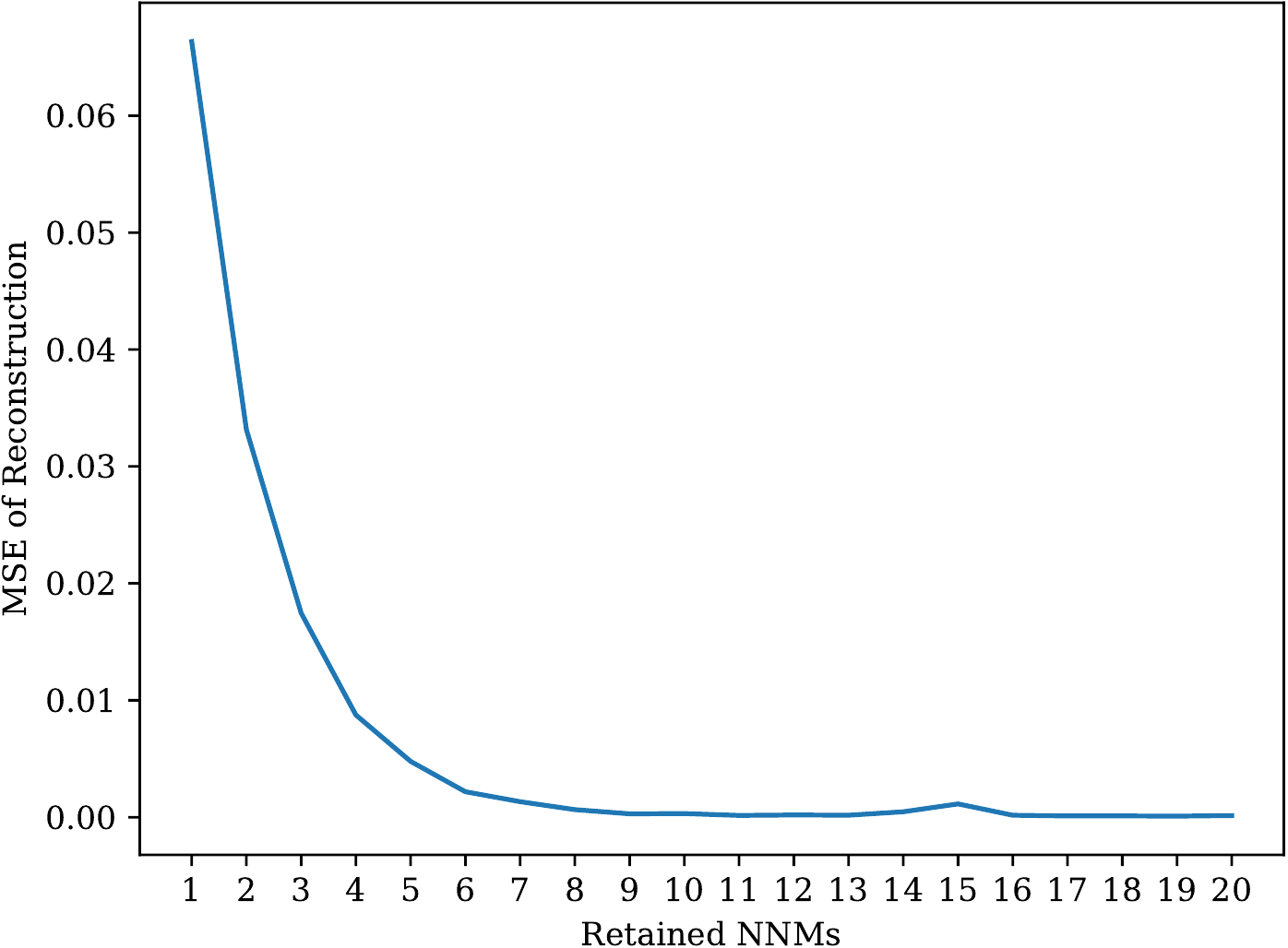}
    \caption{Mean squared error of reconstruction of the autoencoder with varying bottleneck layer sizes for the 20 DOF single nonlinear system.}
    \label{fig:RetainedNNMs}
\end{figure}
8 NNMs were retained for this ROM and the associated autoencoder was accordingly fine tuned. Figure 11 demonstrates the original response and the responses as reconstructed from the 8 retained NNMs for each of the 20 DOF in the system. As observed, a high fidelity to the original response is achieved with this reduced number of retained variables. 

\begin{figure}[h!]
    \centering
   \centerline{ \includegraphics[width=200mm,trim={0cm 0 0mm 0mm},clip]{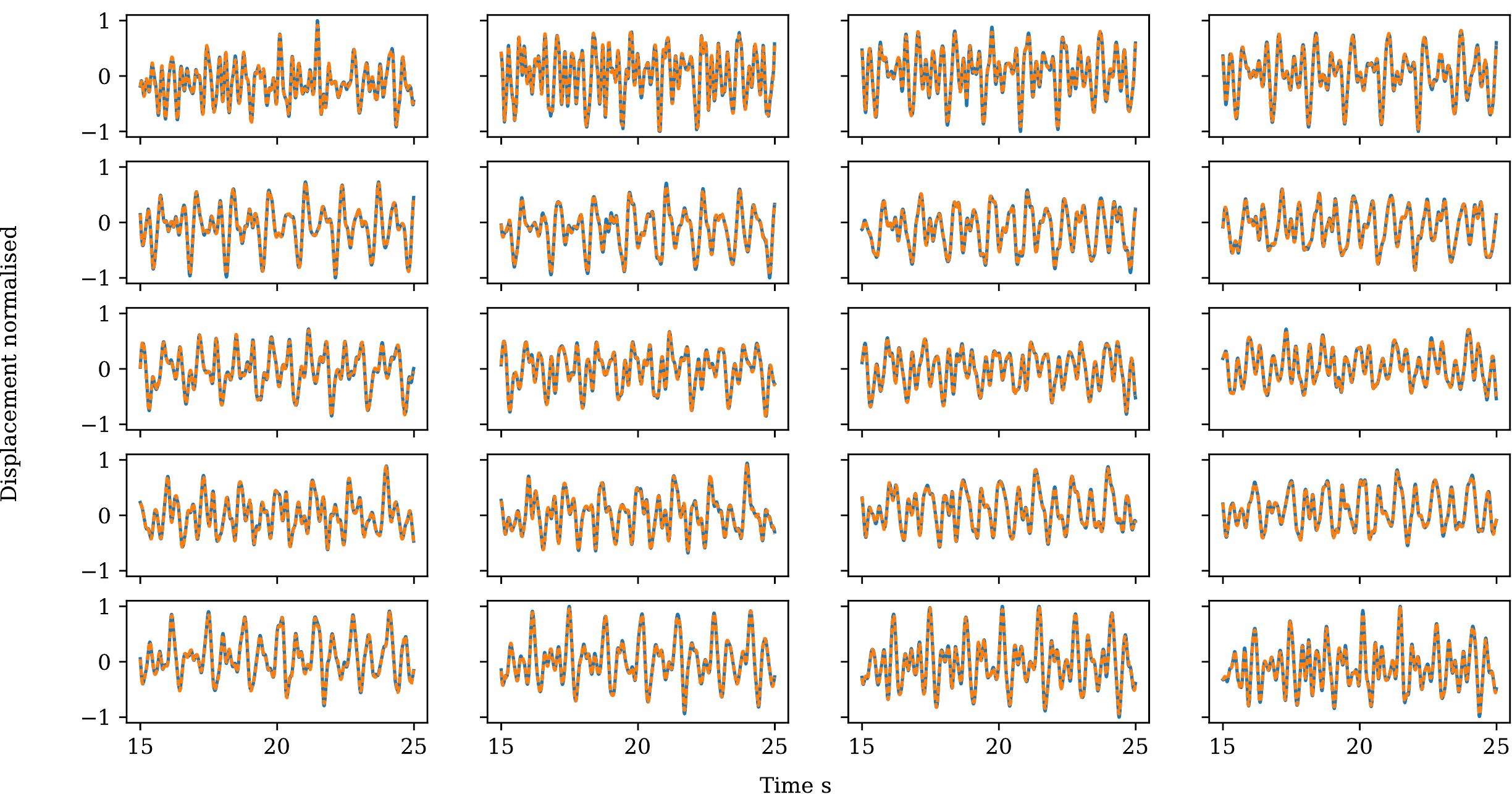}}
    \caption{Comparison of the original displacement time-series (blue solid line) and those reconstructed by an autoencoder comprising 8 nodes in the bottleneck layer (orange dashed line).}
    \label{fig:Recon_1_NL}
\end{figure}

Having trained the autoencoder, the displacement time histories in the physical coordinates are transformed to time histories in the 8 NNMs. These 8 time histories are then used to train a regression model. The regression model hence only has to learn to predict on 8 dimensions as opposed to the initial 20. The final autoencoder used consisted of five layers: a 20 dimensional input layer with linear activation functions, a 20 dimensional hidden layer with \textit{\textit{tanh}} activation functions, an 8 dimensional bottleneck layer with \textit{tanh} activation functions, a 20 dimensional hidden layer with \textit{\textit{tanh}} activation functions followed by a 20 dimensional output layer with linear activation functions. The network was trained using the ADAM optimisation algorithm \cite{kingma2014adam}, on the first 5000 points of the time-series with the testing set on the following 1000 points.The ADAM algorithm is a gradient descent based optimisation method very widely used in the training of neural networks. It tends to exhibit faster convergence than conventional stochastic gradient descent whilst also remaining stable \cite{kingma2014adam}.

\subsubsection{Regression}
An LSTM network was trained, which learns to predict the response of the system in the NNMs given the time-history of the forcing inputs. For the regression, a single \textit{tanh}-activated LSTM unit was used, with a cell state size of 30, with an 8 dimensional fully connected layer as an output layer with linear activation function. The LSTM network was herein setup in an explicitly auto-regressive manner, which is not typically the case. As such, in order to make a prediction at each time step the network input was not only fed the exogenous (Forcing) variables and the hidden state variable from the previous time step, but was further explicitly passed the output (NNM response) values from the previous time steps. In training mode, these previous values were passed in the prediction error manner, meaning that the exact output values of the system at time $t_{k-1}$ were passed on as inputs at time $t_k$. On the testing datasets, however, a simulation error type arrangement was used meaning that the \emph{predicted} outputs at time $t_{k-1}$ were passed as inputs for prediction at time $t_k$. During the prediction phase, the input was formulated again after each prediction to include the autoregressive features, with repeated one step ahead predictions. As such, a stateful LSTM network was used; this simply means that the hidden state values were maintained between each of these single predictions and passed forward to the next prediction \cite{brownlee_2017}. Similarly to the autoencoder, the network was trained on the first 5000 points of the time-series and then predictions were made on the following 1000 points. When training an RNN using the back propagation through time algorithm (BPTT), a maximum number of time steps to back propagate error for each training example must be given. The number of time steps considered is another hyperparameter of the network to be chosen for optimum performance. In this case, 30 time steps were included for the BPTT algorithm. 

\begin{figure}[h!]
    \centering
    \includegraphics[width=160mm]{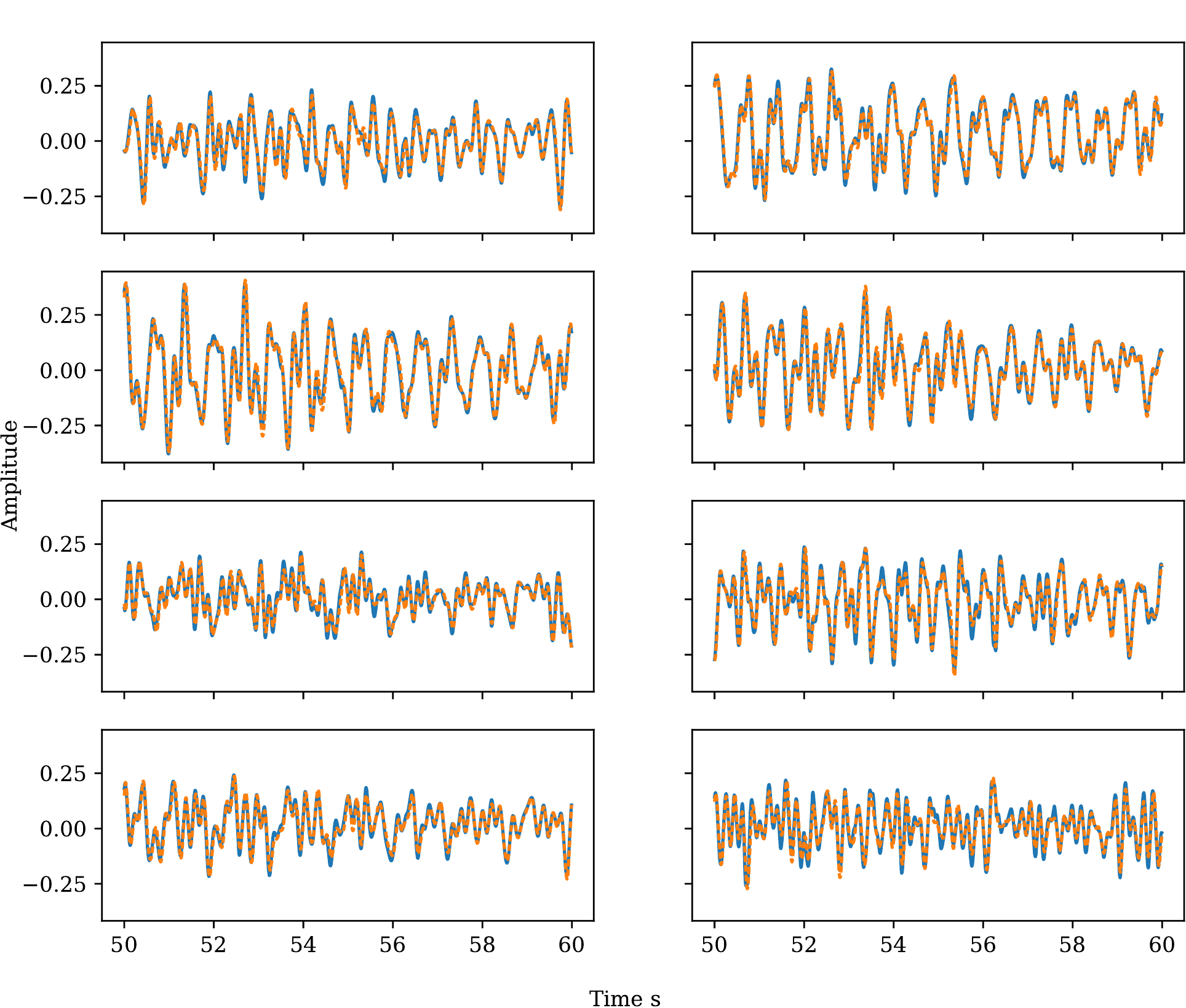}
    \caption{Comparison of the true response (blue line) to predicted response (orange dashed line) in the 8 retained NNMs to a new forcing time history for the 20 DOF single nonlinearity system.}
    \label{fig:Regress_1_NL}
\end{figure}

Figure 12 illustrates the performance of the regression in terms of predicting the response of the system within the 8 retained NNMs. The plot is produced for a new testing input forcing, which was not used during training. The regression demonstrates high fidelity for all of the 8 NNMs and further proves stable, i.e., not tending towards zero or infinity, which might often be a problem with time-series simulations, as a result of accumulating errors.

\subsubsection{Full ROM}
The final step of the ROM procedure pertains to feeding the predicted response within the reduced space through the decoder, which recombines the NNM features to construct the response in the physical coordinate space. Figure 13 shows the result of this decoding, the fidelity of the overall model is shown to be good with a high degree of fidelity between the predicted and true values of the response. It should be noted that the error in the overall model is limited by the fidelity of the compression used in the autoencoder. Even with perfect regression prediction, the error achieved would be that of the reconstruction error of the autoencoder. The prediction of the ROM resulted in a mean squared error, averaged over all 20 DOFs, of $3.2\times10^{-3}$ and a normalised mean squared error, wherein the mean squared error is normalised by the signal power, of $3.1\times10^{-2}$.
\begin{figure}[h]
    \centering
    \centerline{\includegraphics[width=200mm]{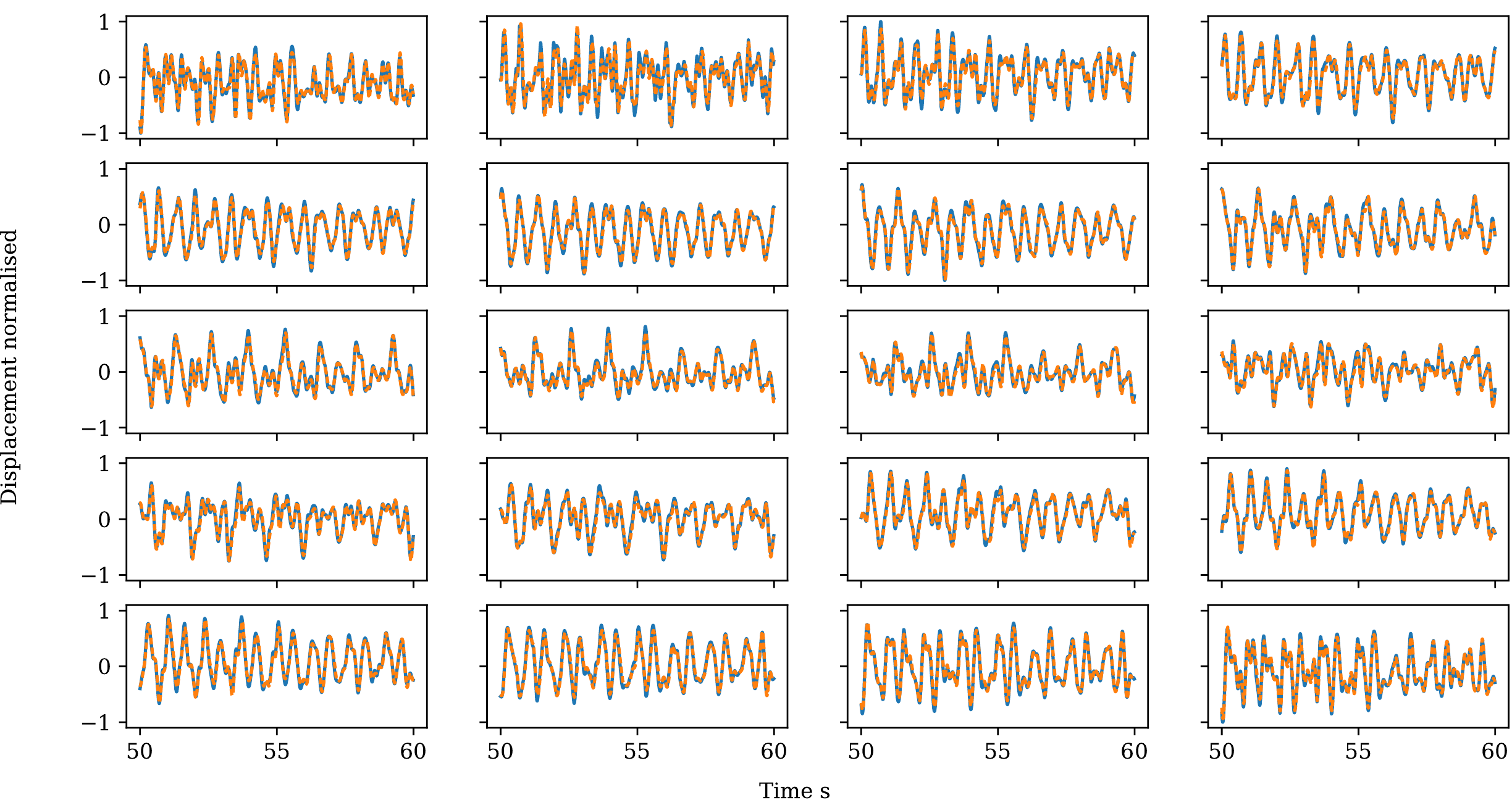}}
    \caption{Comparison of the true response (blue line) to the response predicted by the ROM (orange dashed line) at all DOF for a new forcing time history for the 20 DOF single nonlinearity system.}
    \label{fig:ROM_1_NL}
\end{figure}

\subsubsection{Predicting From Zero Initial State}
The initialisation of the ROM can notably affect its performance. In the previously presented example, we have shown that the ROM can successfully continue prediction form an initiated time-series simulation. It is, however, of interest to evaluate the ROM performance when considering a completely new time-series, i.e., when starting from zero initial conditions. As such, we demonstrate here the prediction of the above trained ROM under utilization of a completely new forcing, assumed to be applied under zero initial conditions. For brevity, only the overall performance of the ROM is here demonstrated. The same autoencoder and LSTM networks, as previously trained, were used for the simulation. Figure 14 compares the ROM prediction to the full order model simulation for a new force, applied at rest (zero initial conditions). The performance of the ROM proves adequate in this respect.

\begin{figure}[h]
    \centering
    \centerline{\includegraphics[width=200mm]{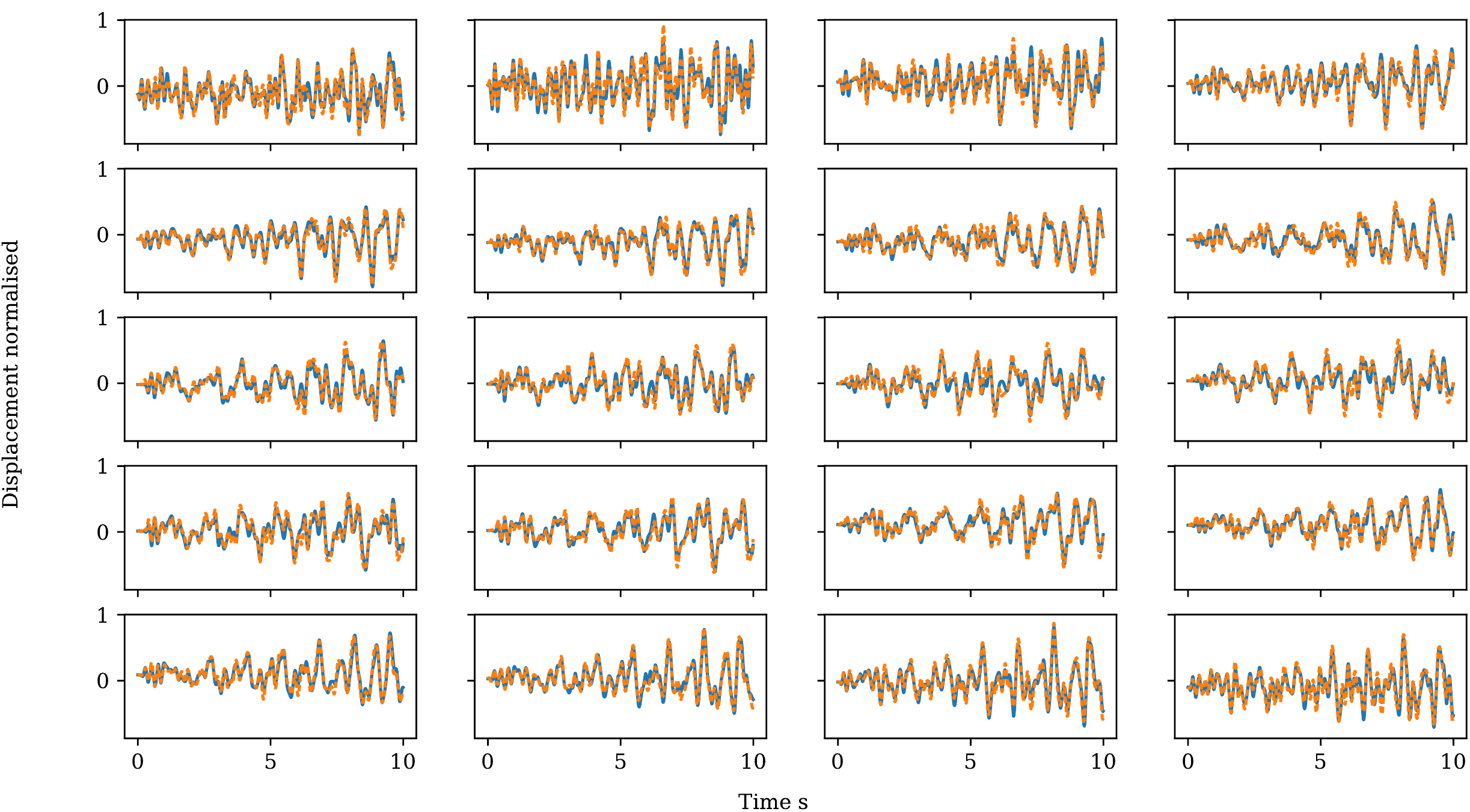}}
    \caption{Comparison of the true response (blue line) to the response predicted by the ROM (orange dashed line) at all DOF for a new forcing time history for the 20 DOF single nonlinearity system when beginning from zero initial condition.}
    \label{fig:From_zero_rom}
\end{figure}

However, it should be noted that the training has to be carried out in a way that ensures inclusion of zero initial conditions in the training dataset. An LSTM network, which has a lookback value of $n$, naturally requires availability of the previous $n$ datapoints to predict the value at $n+1$. As such, when creating a training dataset, it is common to omit the first $n$ output values. If, however, it is desired that the LSTM network is trained to predict from zero initial conditions, which is typically desired for a metamodel, a possible solution lies in padding the beginning of the time-series with zeros.

\subsection{20 Nonlinear DOFs}
In extending to a more complex case, the 20 DOF system described previously is augmented by inclusion of cubic nonlinear spring elements between each of the DOFs, in addition to the linear elements as shown in Figure 15. The parameters of each element remain unchanged from the previous example. 

\begin{figure}[h!]
    \centering
    \includegraphics[width=150mm]{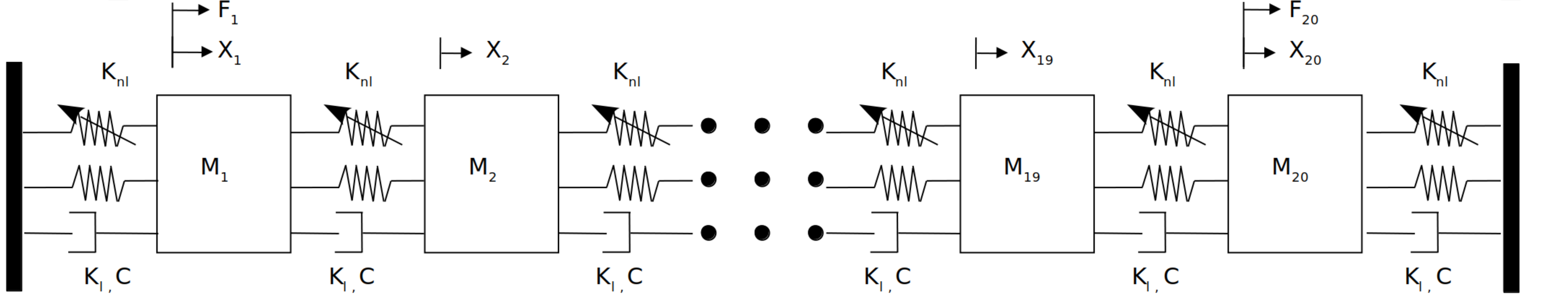}
    \caption{20 Degree of freedom mass spring damper system with nonlinearities between each of the masses.}
    \label{fig:20DOF_20NL}
\end{figure}

Forcing was applied to the system at each of the end DOF as a low-pass filtered white noise signal with different realisations of the signal applied at either end. Each signal had a variance of $2.5 N^2$. The cut-off frequency of the low pass filter was set at $7.5 Hz$ which lies between the 11th and 12th natural frequencies of the linearised system. The time-series solution of the system was performed using the 4th order Runge-Kutta integration scheme at a time step of $0.01 s$. Figure 16 shows the magnitude of the restoring force in the linear and nonlinear spring elements between the 10th and 11th DOFs in the system. Figure 17 plots the total restoring force against displacement between the 10th and 11th DOFs. The restoring force demonstrates a clear strain hardening nonlinear behaviour. Table \ref{tab:01} presents the ratio of the root mean squared value of the nonlinear to the linear restoring forces at each of the DOFs. This demonstrates that a similar level of nonlinearity is present at every DOF and hence this constitutes a highly complex nonlinear system.

\begin{figure}[h!]
    \centering
    \includegraphics[width=170mm]{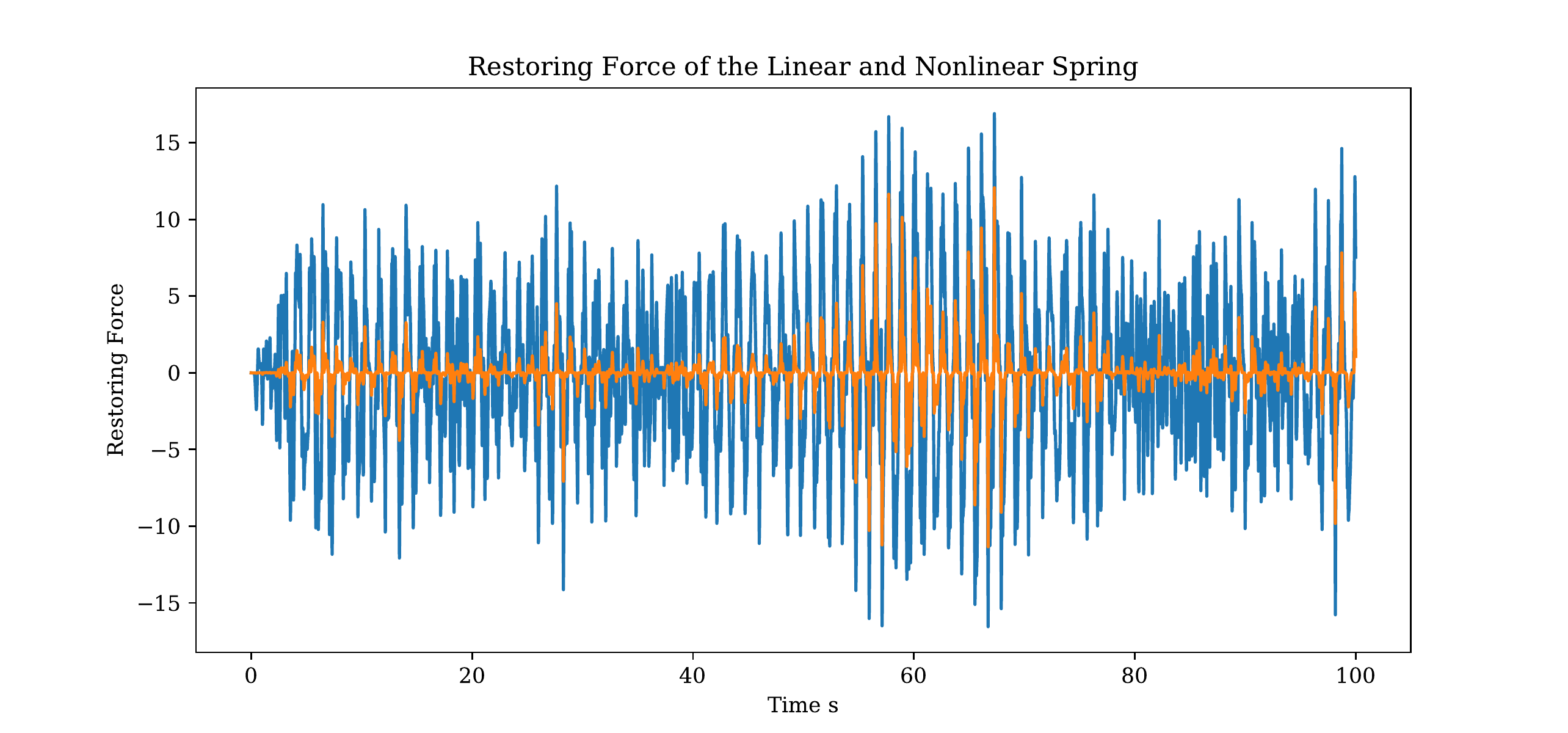}
    \caption{Relative magnitude of the linear and nonlinear restoring forces at the tenth degree of freedom.}
    \label{fig:RestForce_20NL}
\end{figure}

\begin{figure}[H]
    \centering
    \includegraphics[width=90mm]{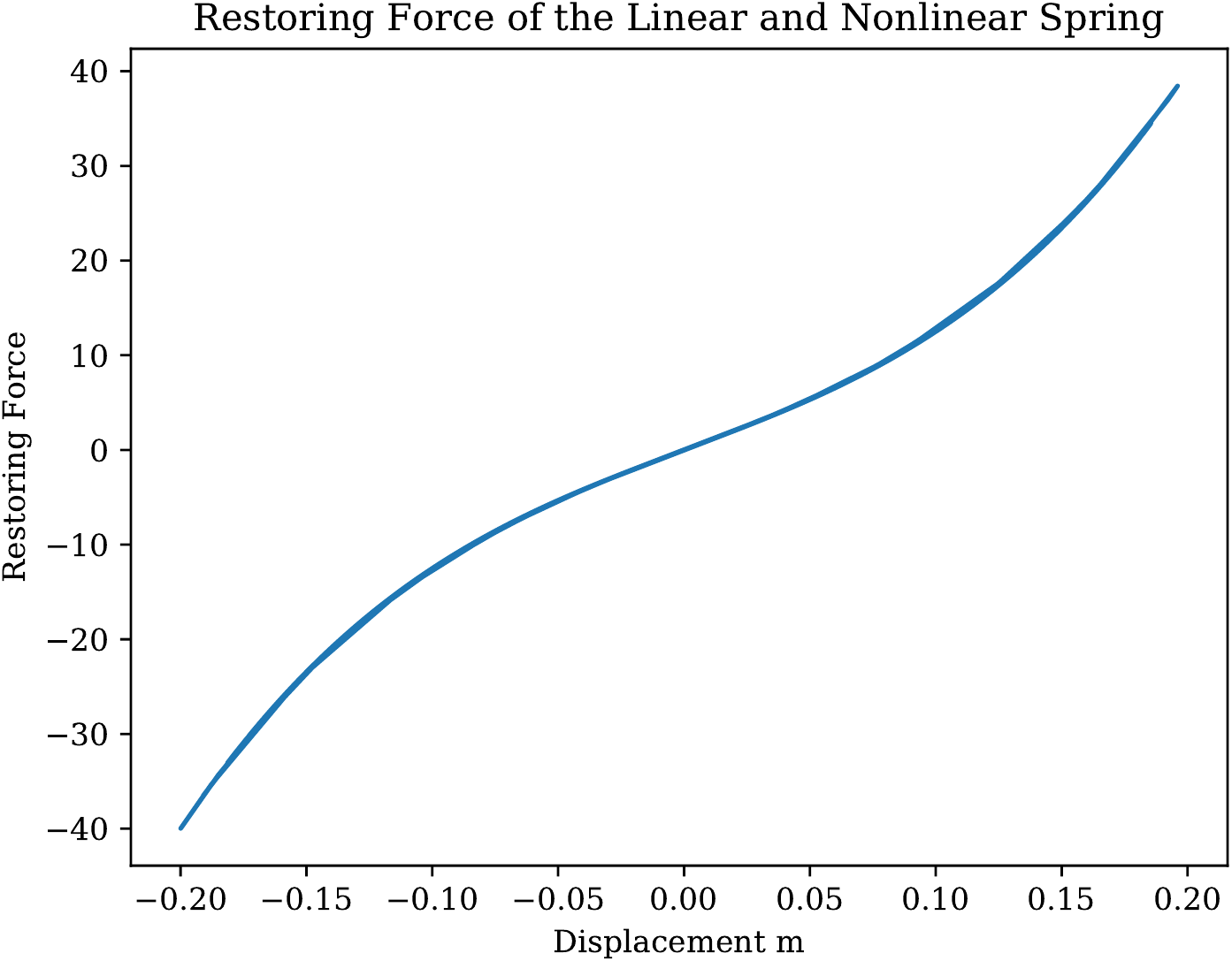}
    \caption{Force-Displacement plot of the restoring force between the 10th and 11th degrees of freedom.}
    \label{fig:my_label}
\end{figure}

\begin{table}[H]

\centering
\caption{Ratio of RMS of the linear and nonlinear restoring forces at each of the DOF for the 20 DOF multiple nonlinear system.}
\begin{tabular}{lcccccccccc} 
\hline
DOF       & 1                         & 2                         & 3                         & 4                         & 5                         & 6                         & 7                         & 8                         & 9                         & 10                         \\
RMS Ratio & \multicolumn{1}{l}{0.144} & \multicolumn{1}{l}{0.183} & \multicolumn{1}{l}{0.198} & \multicolumn{1}{l}{0.215} & \multicolumn{1}{l}{0.230} & \multicolumn{1}{l}{0.235} & \multicolumn{1}{l}{0.240} & \multicolumn{1}{l}{0.244} & \multicolumn{1}{l}{0.247} & \multicolumn{1}{l}{0.250}  \\ 
\hline
DOF       & 11                        & 12                        & 13                        & 14                        & 15                        & 16                        & 17                        & 18                        & 19                        & 20                         \\
RMS Ratio & \multicolumn{1}{l}{0.250} & \multicolumn{1}{l}{0.247} & \multicolumn{1}{l}{0.244} & \multicolumn{1}{l}{0.240} & \multicolumn{1}{l}{0.235} & \multicolumn{1}{l}{0.230} & \multicolumn{1}{l}{0.216} & \multicolumn{1}{l}{0.200} & \multicolumn{1}{l}{0.187} & \multicolumn{1}{l}{0.147} 
\end{tabular}

\label{tab:01}
\end{table}

\subsubsection{Reconstruction}
Similarly to the previous case, autoencoders were trained with varying numbers of retained NNMs. Figure 18 shows the plot of the reconstruction error against the number of retained NNMs. It can once again be seen that after a given number of retained dimensions, the error of reconstruction asymptotically approaches zero. For this ROM it was decided to retain 9 NNMs, as always the number of retained NNMs can be selected when creating the ROM to balance fidelity and reduction.

\begin{figure}[h!]
    \centering
    \includegraphics[width=100mm]{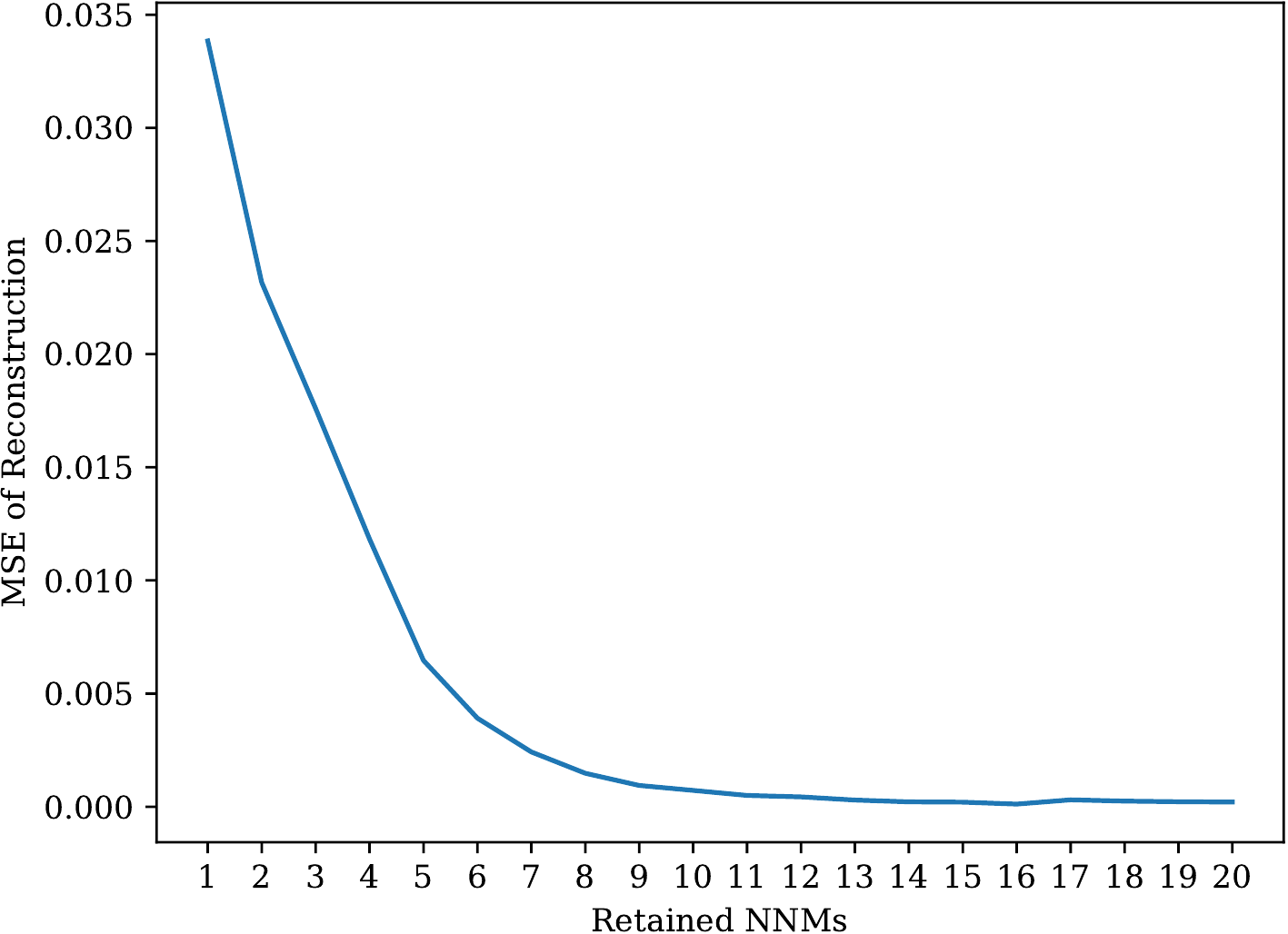}
    \caption{Mean squared error of reconstruction of the autoencoder with varying bottleneck layer sizes for the 20 DOF multiple nonlinear system.}
    \label{fig:20NLRetained}
\end{figure}

For this system, 9 NNMs were retained and, consequently, an autoencoder with 9 hidden units in the bottleneck layer was trained. The adopted autoencoder consisted of five layers: a 20 dimensional input layer with linear activation functions, a 20 dimensional hidden layer with \textit{tanh} activation functions, a 9 dimensional bottleneck layer with \textit{tanh} activation functions, a 20 dimensional hidden layer with \textit{tanh} activation functions followed by a 20 dimensional output layer with linear activation functions. The network was trained using the ADAM optimisation algorithm. The network was trained on the first 5000 points of the time-series with the testing set on the following 1000 points. Figure 19 shows the fidelity of this reduction via a comparison between the reconstructed response and the original response time-series.

\begin{figure}[h!]
    \centering
   \centerline{ \includegraphics[width=200mm]{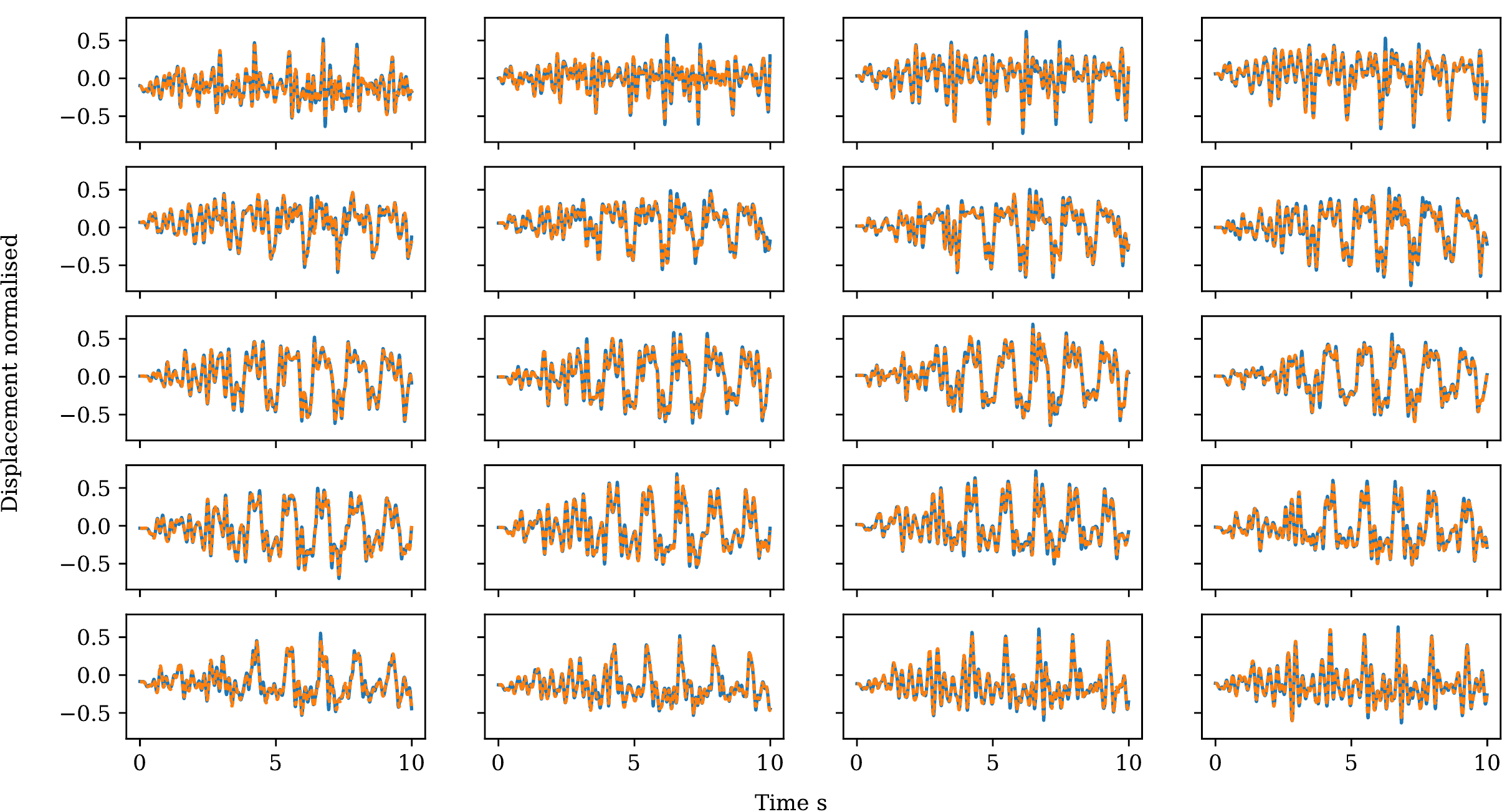}}
    \caption{Comparison of the original displacement time-series (blue solid line) and those reconstructed by an autoencoder comprising 9 nodes in the bottleneck layer (orange dashed line).}
    \label{fig:Recon_20_NL}
\end{figure}

\subsubsection{Regression}
The design of a network architecture for this regression problem was found to be notably more difficult than in the previous case. The size of the cell state required to adequately approximate the system via a single layer LSTM network resulted in a high number of parameters. This is presumably as a result of the highly nonlinear nature of the system requiring a more complex network for an adequate approximation. As such, a 2 layer stacked LSTM network was also trialled, which was found to achieve improved results in comparison to a single layer network with far fewer parameters. The higher efficacy of deeper networks in representing complex functions than wide networks is considered to be a standard result in machine learning \cite{BengioLecun}. The adopted network architecture comprises a \textit{tanh}-activated LSTM unit with a cell state size of 27, stacked with another \textit{tanh} activated LSTM unit with a cell state size of 20 and finally a fully connected output layer of dimension 9 with a linear activation. Similarly to previous examples, the first 5000 time points were used for training with the next 1000 used for testing. For the predictive network a stateful LSTM network was once again used and the number of points considered in the BPTT algorithm was 50.

\begin{figure}[h!]
    \centering
    \centerline{\includegraphics[width=190mm]{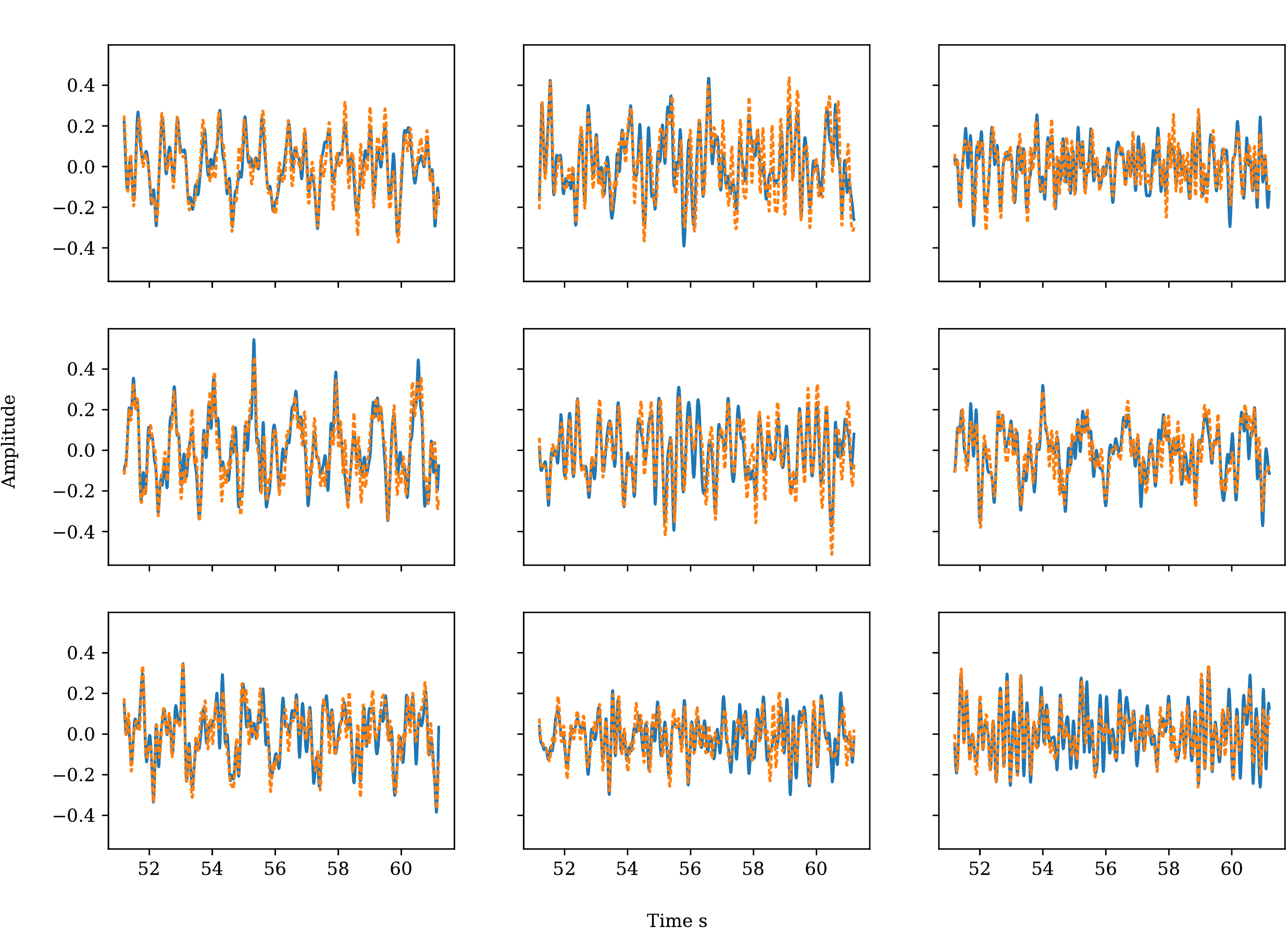}}
    \caption{Comparison of the true response (blue line) to predicted response (orange dashed line) in the 9 retained NNMs to a new forcing time history for the 20 DOF multiple nonlinearity system.}
    \label{fig:Regress_20_NL}
\end{figure}

Figure 20 shows the performance of the regression in predicting the response of the system within the 9 retained NNMs for a new testing input forcing which was not used during training. The regression demonstrates reasonable fidelity for all of the 9 NNMs, however, with notably decrease performance compared to the previous example. It is again noteworthy that the regression provides stable predictions without divergence. This is notable as the regression was trained in \say{prediction error} mode which in the case of NARX models can cause issues with simulation stability.

\subsubsection{Full ROM}
As previoulsy mentioned, in a final step the predicted response in the lower dimensional space (NNMs) is passed through the decoder, which recovers the full order response in the physical coordinate space. Figure 21 shows the result of this decoding. The overall performance of the ROM is fairly successful, especially in those DOFs dominated by the lower frequency NNMs. In modes exhibiting higher frequency content, however, there is some loss in fidelity from the ROM. Some of this loss in fidelity can naturally be prescribed to the limited number of retained NNMs. As later demonstrated, in section: Physical Interpretation of NNMs, the addition of more nodes in the hidden layer tends to increase the amount of high frequency content captured. The overall performance of the ROM is also shown to be stable over the 10 seconds of simulation. The prediction of the ROM resulted in a mean squared error, averaged over all 20 DOFs, of $7.5\times10^{-3}$ and a normalised mean squared error, wherein the mean squared error is normalised by the signal power, of $1.51\times10^{-1}$.

\begin{figure}[h!]
    \centering
    \centerline{\includegraphics[width=200mm]{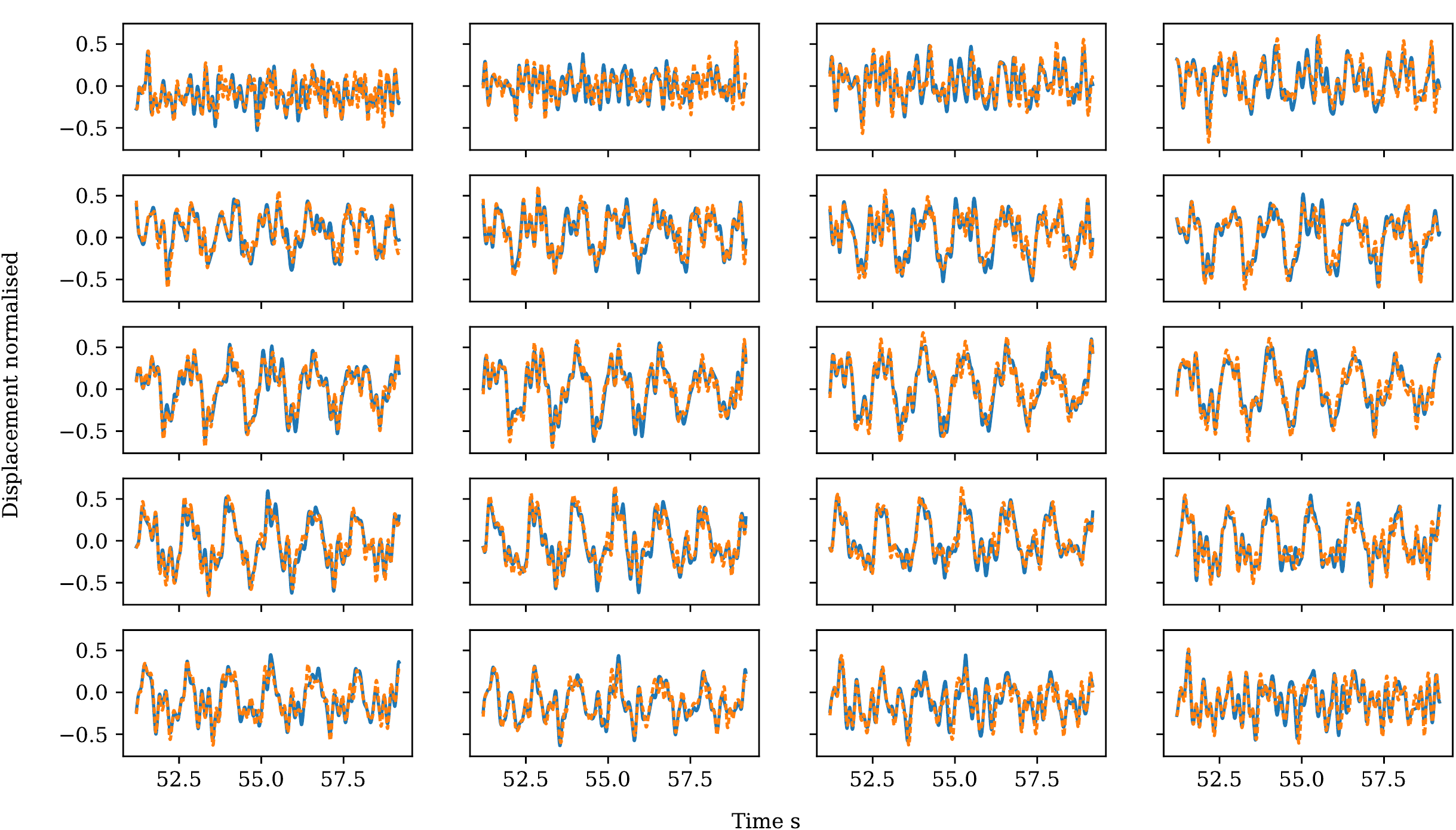}}
    \caption{Comparison of the true response (blue line) to the response predicted by the ROM (orange dashed line) at all DOF for a new forcing time history for the 20 DOF multiple nonlinearity system.}
    \label{fig:ROM_20_NL}
\end{figure}

\subsection{Bouc-Wen Frame Structure}
To further test the reduction framework, a more realistic system is finally analysed. This numerical case study simulates a 3D two story building modeled with nonlinear links exhibiting a Bouc-Wen type nonlinearity. For the set up of this frame structure, all beams and columns are of a rectangular cross section ($40\: cm  \times 40\: cm$) and are considered to be constructed of reinforced concrete. The frame is considered only under self-weight, without addition of a further mass. The material parameters assumed are: Young modulus $E = 21\: GPa$, Poissons' ratio $\nu = 0.20$ and density $\rho = 2400\: kgm^{-3}$. The length of the frame is $l = 7.5\:m$, the width is $w = 5\:m$ and the height $h = 3.2\:m$. Each node has six degrees of freedom, three translations and three rotations. This gives the frame structure a total of 108 degrees of freedom. This example system is based on that demonstrated in Vlachas et al. \cite{VlachasK/nme.6447}.

Bouc-Wen type hysteretic links are added to the node-to-node connections of the frame, as additional elements. These links produce a hysteretic internal forcing dependent on the relative displacement between the two connecting nodes. This is the case for all element connections within the frame structure, as well as for the frame supports, which are considered to be connected to the ground via additional Bouc-Wen hysteretic links. The frame is mass-proportionally damped with 4 \% damping applied in the first mode. The geometry of the frame is demonstrated in Figure 22 below. The Bouc-Wen model is a flexible representation of hysteresis. The theoretical background of the Bouc-Wen model along with a simulated system identification problem is given in \cite{Chatzi2009}, whilst application of the Bouc-Wen model for identification of physical structures is given in \cite{Chatzi2010,Miah_2015}.

\begin{figure}[h!]
    \centering
    \includegraphics[width=110mm]{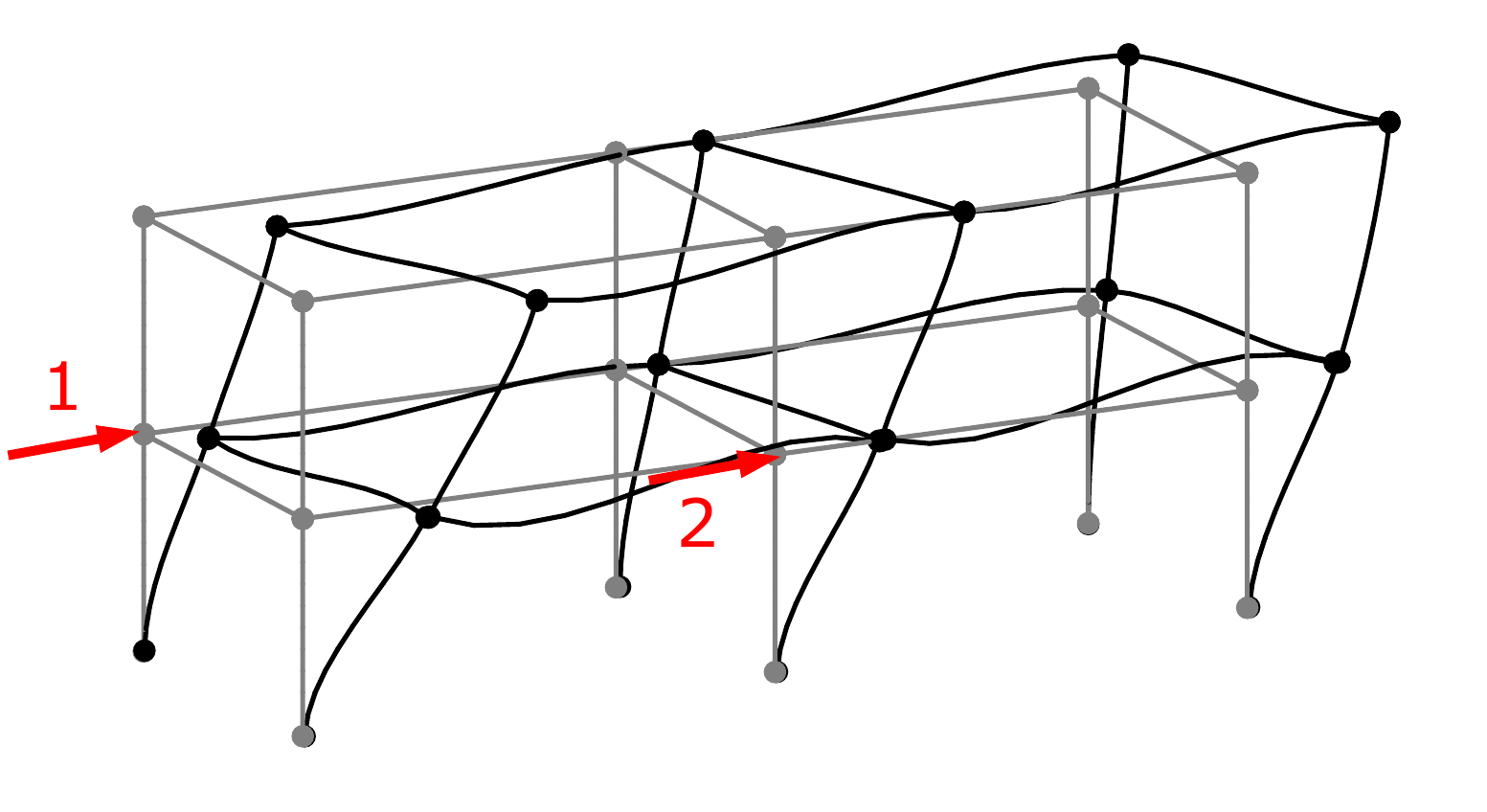}
    \caption{108 Degree of freedom frame structure with hysteretic links between each node modelled with Bouc-Wen nonlinearity. Force application points indicated by the red arrows.}
    \label{fig:BWframe}
\end{figure}

A Bouc-Wen model is used for describing the behaviour of the hysteretic links, as described in equation \ref{eq:BW}: wherein $x$ denotes the displacement of the link, $\mathbf{z}$ the hysteretic parameter and $A$ controls the hysteresis amplitude. Parameter $w$ along with the hysteretic parameters $\beta$ and $\gamma$ determine the basic shape of the hysteresis loop. Their absolute values are not of interest, but rather their sum/difference that may define a hardening or softening behaviour. In this case, the Bouc-Wen model is parameterised only in terms of $\textbf{A}$ and $\textbf{z}_{max}$ for which values of $0.3$ and $1e-04$ respectively are used.

\begin{equation}
    \label{eq:BW}
    \begin{aligned}
    \mathbf{ \dot{z}} = \mathbf{A}\dot{x}-\beta\lvert\dot{x}\rvert\mathbf{z}\lvert\mathbf{z}\rvert^{n-1} - \gamma \dot{x}\lvert\mathbf{z}\rvert^n \\ \mathbf{z}_{max} = \Big(\frac{\mathbf{A}}{\beta + \gamma}\Big)^{\frac{1}{w}}
    \end{aligned}
\end{equation}

Forcing is applied in the form of two asymmetric sinusoidal loads of different amplitude at $1.5 Hz$. This forcing frequency lies between the 2nd and 3rd natural frequencies of the linearised system. The location and direction of the applied forces are indicated in Figure 22. The force applied at location 1 features an amplitude of $30 kN$, while the forcing at location 2 has an amplitude of $60 kN$. Since for the frame structure, a harmonic forcing was used rather than random excitation, a different training and testing regime was selected to assess the performance of the ROM method. For training of the ROM, 3 different simulations were generated with different amplitudes of the harmonic forcing. A testing simulation was then created using a 4th amplitude of forcing which was then used to assess whether the ROM method could generalise to different forcing amplitudes. For each of the different amplitudes 500 time steps were simulated with an integration step of $0.01 s$ and a nonlinear implementation of the Newmark integration scheme was used \cite{chopra_2012}. Training datasets were generated with forcing amplitudes set at $1, 0.5, 0.25$ whilst testing was performed with a forcing amplitude of $0.75$. The simulated frame exhibits differing amounts of nonlinearity in the response in the different hysteretic links. The links exhibiting greater amplitudes of displacement, those on the upper story, are subjected to a considerably larger amount of nonlinearity, whilst in some of the links in the lower story, no nonlinearity is observed in  the response. Figure 23 shows the restoring force plotted against displacement for 2 of the hysteretic links in the Bouc-Wen frame structure. Figure 23a is taken from an upper story link exhibiting relatively large displacements and correspondingly exhibiting a clear nonlinear hysteretic behaviour. Figure 23b, however, is taken from a lower story link with much lower displacement level and correspondingly very low amounts of nonlinearity in the restoring force. The majority of the links in the frame structure exhibited a level of nonlinearity between that of these two examples.

\begin{figure}[h!]
\centering
\begin{subfigure}{.5\textwidth}
  \centering
  \includegraphics[width=80mm,trim={0mm 0mm 0mm 0mm},clip]{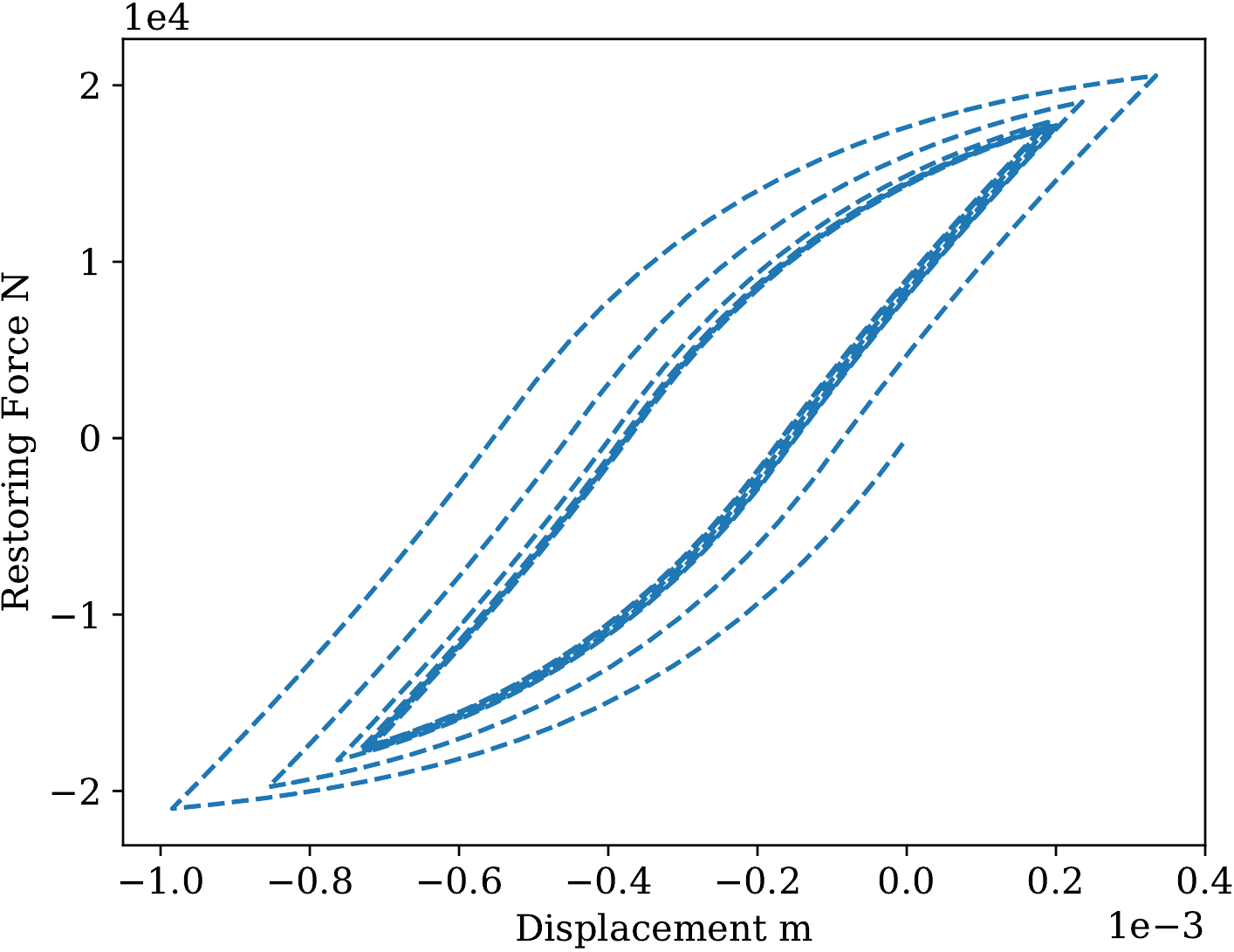}
  \caption{Restoring force of an upper story link.}
  \label{fig:sub1}
\end{subfigure}%
\begin{subfigure}{.5\textwidth}
  \centering
  \includegraphics[width=80mm,trim={0mm 0mm 0mm 0mm},clip]{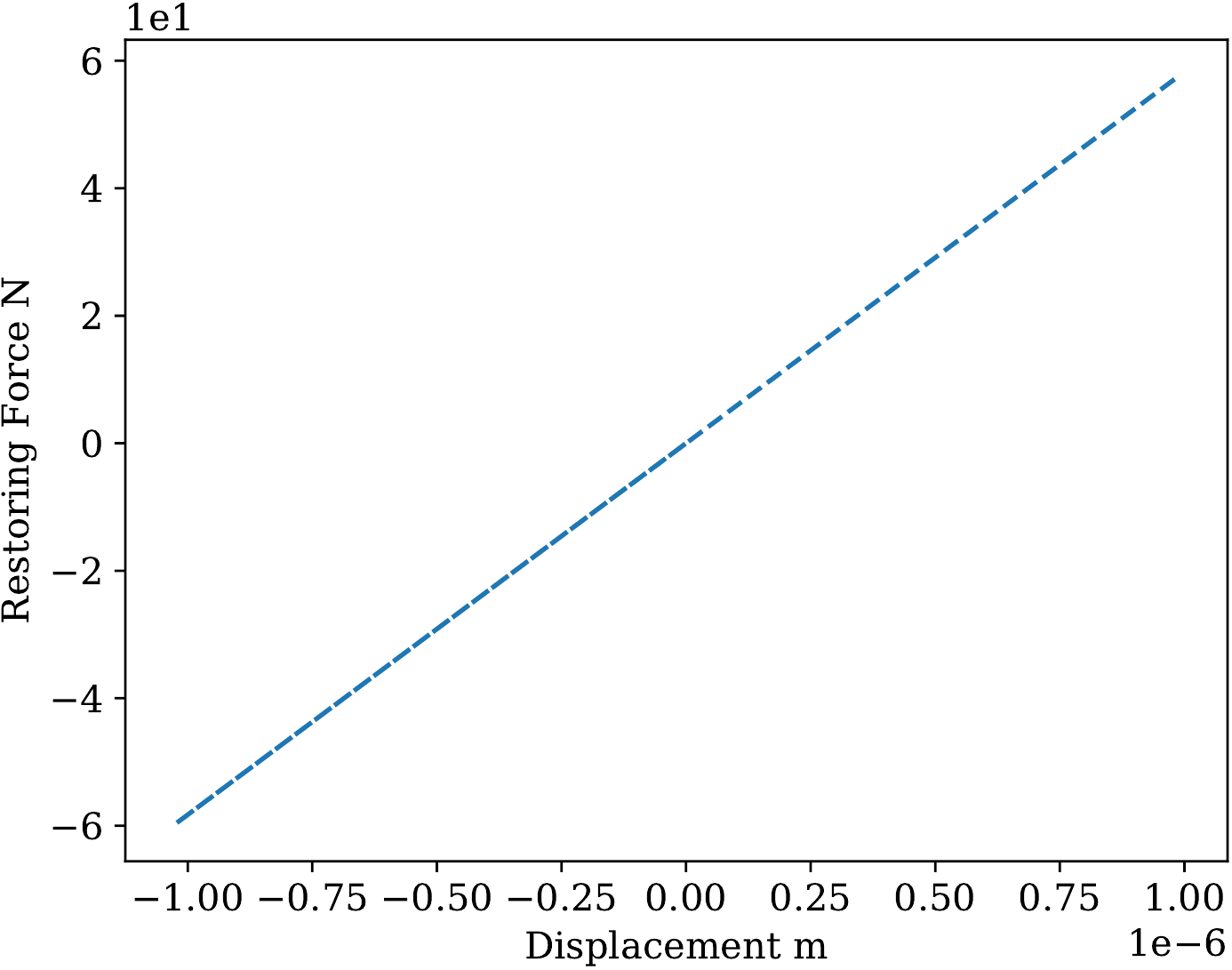}
  \caption{Restoring force of a lower story link.}
  \label{fig:sub2}
\end{subfigure}
\caption{Restoring force of hysteretic links in the Bouc-Wen frame structure.}
\label{fig:BW_Rest_force}
\end{figure}

\subsubsection{Reconstruction}
Similarly to the previous cases, autoencoders were trained with varying numbers of retained NNMs. Figure 24 shows the plot of reconstruction error against the number of retained NNMs. In this case a log scale was used to better visualise the error. In this case, the error very quickly dropped to a very low value allowing for high fidelity with only few retained NNMs, herein it was decided to retain 6 NNMs. The final autoencoder used consisted of five layers: a 108 dimensional input layer with linear activation functions, a 108 dimensional hidden layer with \textit{tanh} activation functions, a 6 dimensional bottleneck layer with \textit{tanh} activation functions, a 108 dimensional hidden layer with \textit{tanh} activation functions followed by a 108 dimensional output layer with linear activation functions. The training set of this autoencoder consisted of the displacement time histories from each of the 3 training simulations carried out. Since the autoencoder trained herein was a static autoencoder, this allowed these training datasets to be considered simultaneously during network training without difficulty.

\begin{figure}[h!]
    \centering
    \includegraphics[width=100mm,trim={0mm 0mm 0mm 0mm},clip]{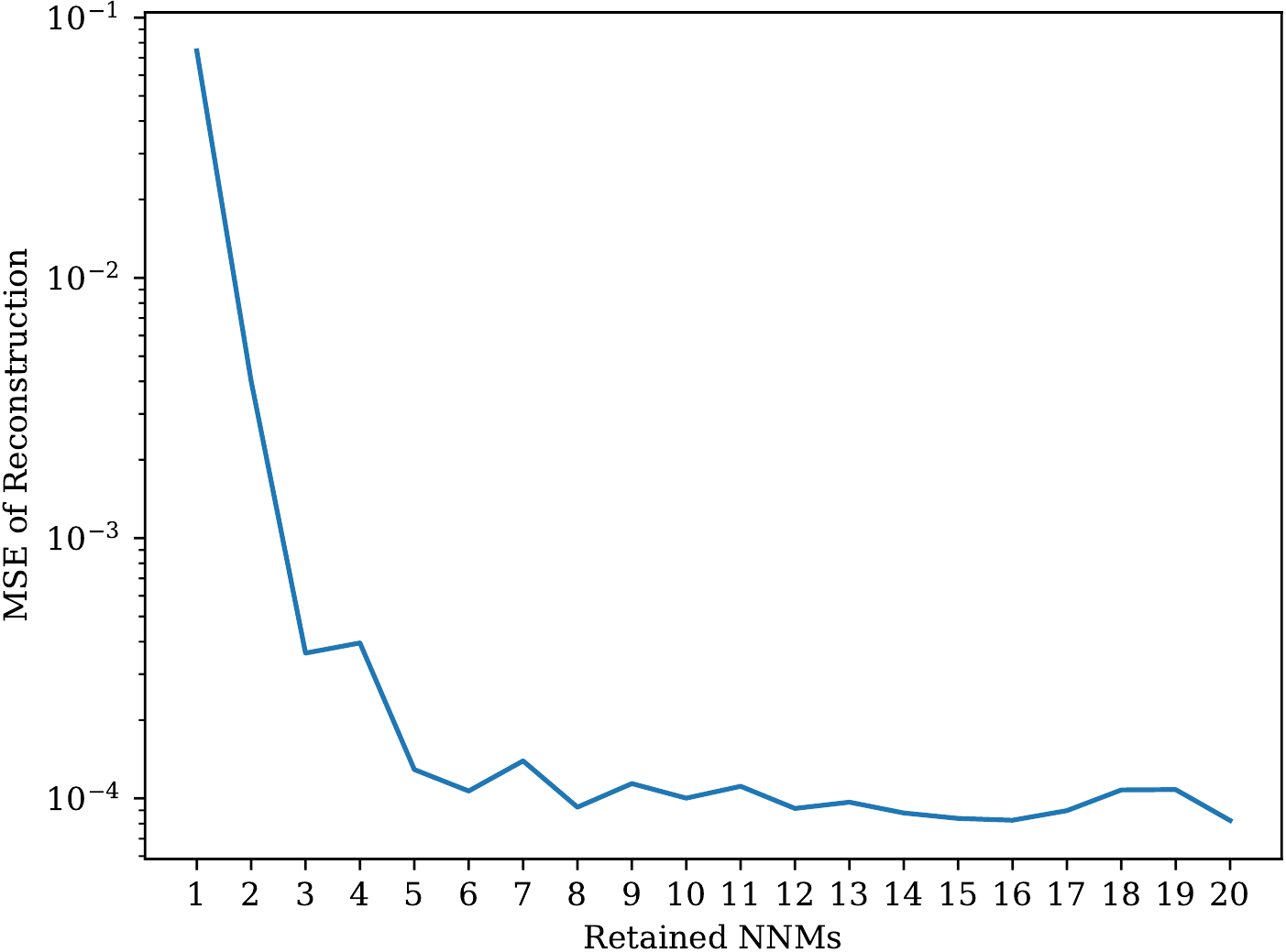}
    \caption{Mean squared error of reconstruction of the autoencoder with varying bottleneck layer sizes for the Bouc-Wen frame structure.}
    \label{fig:RetainedNNMs_BW}
\end{figure}

Figure 25 demonstrates the fidelity of the reconstruction provided by the autoencoder on the basis of 6 retained NNMs, in comparison to the original displacement time-series of the response at the physical DOF. Results are presented for 20 of the in total 108 DOF for brevity, these 20 presented DOF are from various locations and orientations on the structure. The results at the remaining DOFs, not presented in Figure 25, were of similar fidelity. The results presented here are the performance of the autoencoder on the testing dataset, that generated at a different forcing amplitude to the training datasets. The achieved high fidelity demonstrates that not only can the full response be accurately recovered from the 6 NNMs, but also that the employed transform can also generalise to different amplitudes of forcing.

\begin{figure}[h!]
    \centering
   \centerline{ \includegraphics[width=200mm,trim={0cm 0 0mm 0mm},clip]{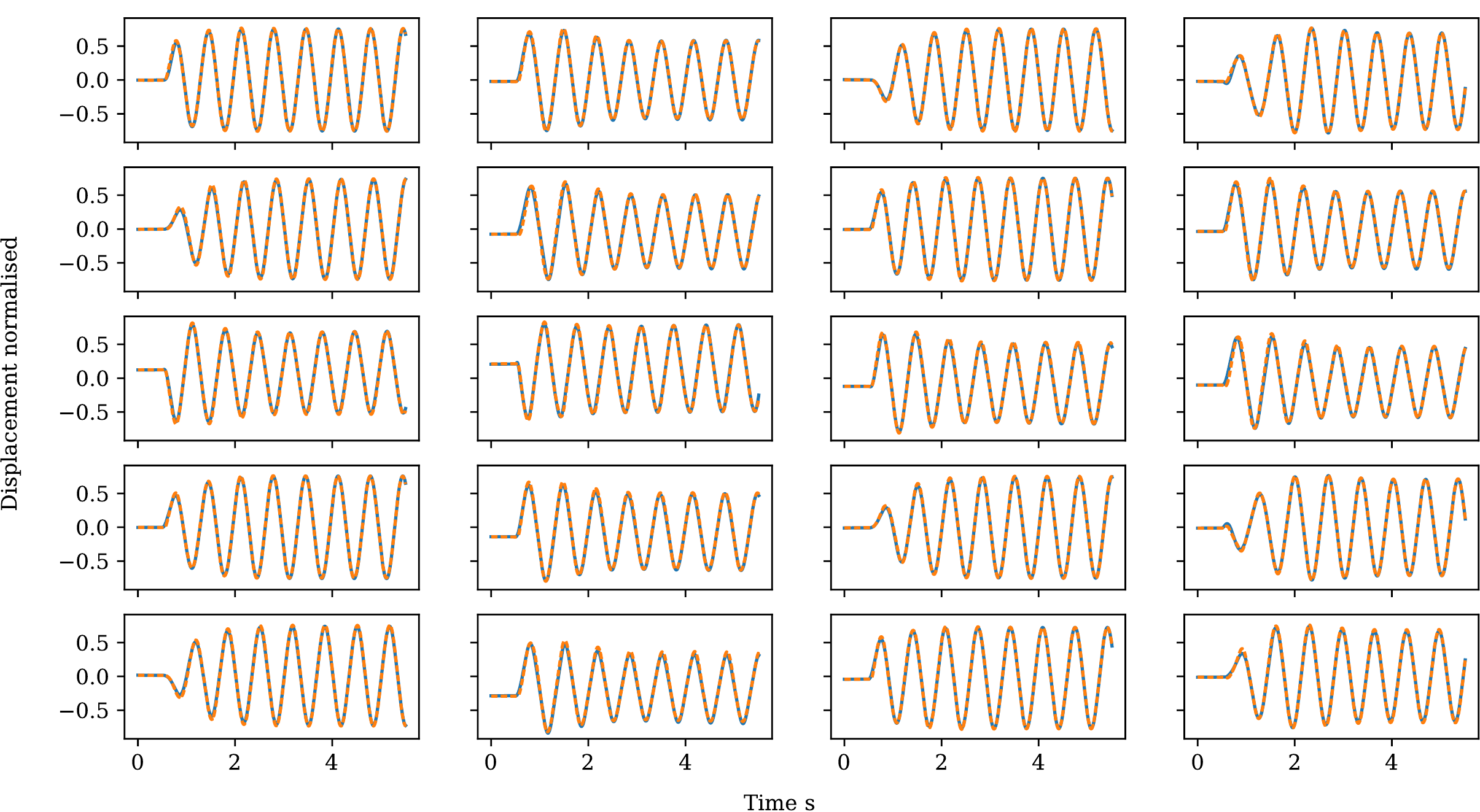}}
    \caption{Comparison of the original displacement time-series (blue solid line) and those reconstructed by an autoencoder comprising 6 nodes in the bottleneck layer (orange dashed line).}
    \label{fig:Recon_BW}
\end{figure}

\subsubsection{Regression}

When training an LSTM network for this use case, special care must be taken with regards to the training batch size. In order to account for amplitude dependencies within the system, 3 different training time-series were generated at different amplitudes. However, unlike in the case of the random forcing, these are 3 non-contiguous time-series. As such, when training the LSTM network it is important that the internal state of the network is reset to zero between training on each of the three time-series. In most common implementations of LSTM networks, the resetting of the cell state of the network can be controlled by the user. For multiple non contiguous time-series of the same length (as is here the case), the easiest way to do this is to have the cell state reset after each training batch and to have the training batch size equal to the length of each of the individual time-series.

Having considered this particular detail regarding the network state, the training of the LSTM network in this test case proved less challenging than in the previous cases. A single layer network with a relatively small hidden state size was found to be sufficient. The final network used in this case consisted of a \textit{tanh} activated LSTM network layer with 15 cell state units, followed by a linear activated fully connected output layer with 6 units. The network was trained using the three training time-series each of 500 points and tested on the testing time-series also of 500 points. For the predictive network, a stateful LSTM network was used and for training using the BPTT algorithm 20 time steps were considered.

Figure 26 demonstrates the performance at predicting the system response in the 6 NNMs for the testing dataset at a different forcing amplitude. Performance of the network is high especially in the steady state response. In this case as a zero initial state was used, the transients in the system do have some effect hence the slightly worse performance initially.

\begin{figure}[h!]
    \centering
    \centerline{\includegraphics[width=190mm,trim={0cm 0 0mm 0mm},clip]{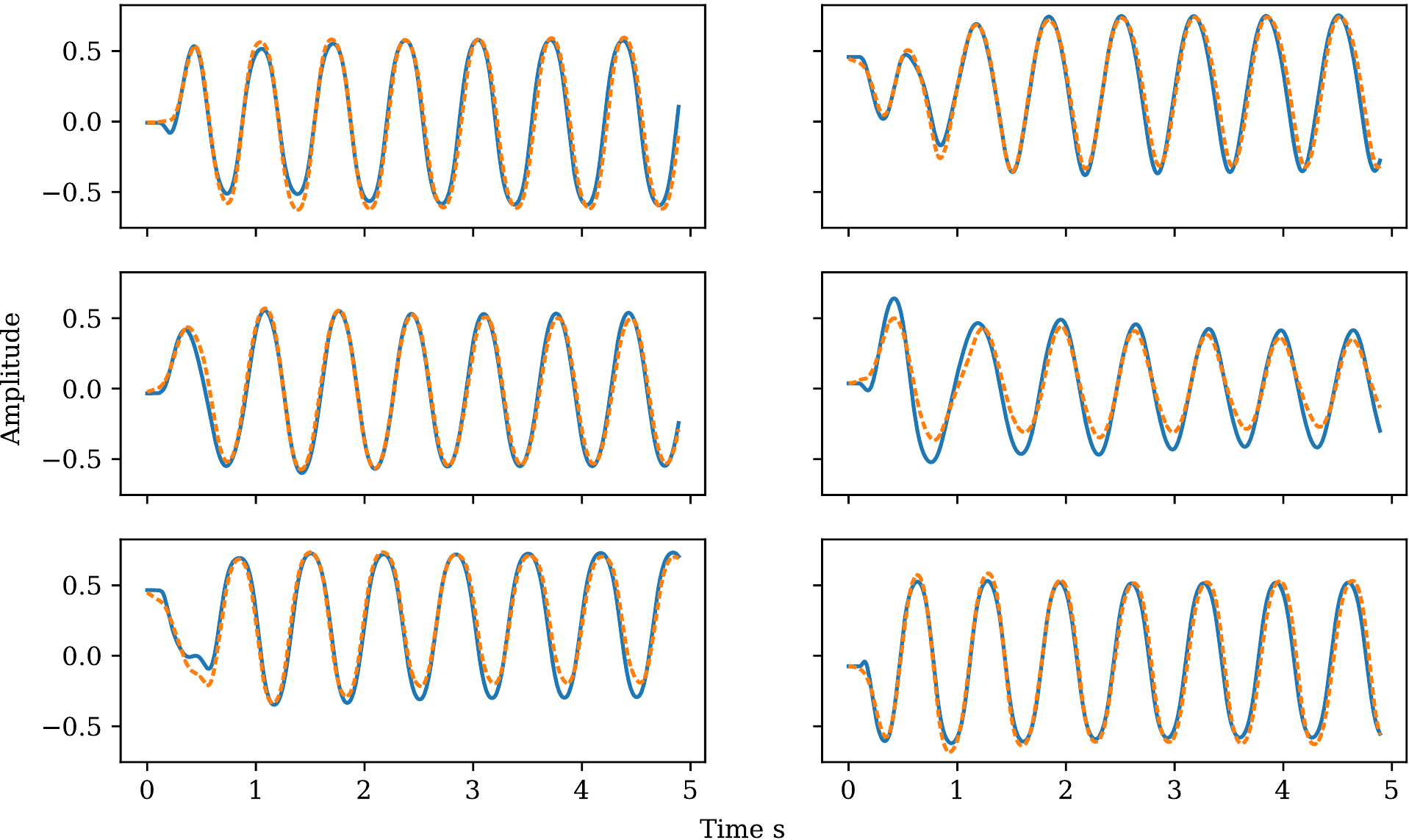}}
    \caption{Comparison of the true response (blue line) to predicted response (orange dashed line) in the 6 retained NNMs to a new forcing time history for the 108 DOF Bouc-Wen frame system.}
    \label{fig:Regress_BW}
\end{figure}

\subsubsection{Full ROM}

The final stage of the ROM procedure is to pass the predicted response within the NNMs through the decoder which recombines them to construct the response in the physical coordinate space. Figure 27 shows the result of this decoding. The overall performance of the ROM is of high standard with some slight differences in phase and additional loss of fidelity during the transient phase. The overall response is demonstrated here in 20 of the in total 108 DOF of the system but are representative of the results achieved in all DOF of the system. The prediction of the ROM resulted in a mean squared error, averaged over all 20 DOFs, of $4.5\times10^{-4}$ and a normalised mean squared error, wherein the mean squared error is normalised by the signal power, of $2.1\times10^{-2}$.

\begin{figure}[H]
    \centering
    \centerline{\includegraphics[width=200mm,trim={0cm 0 0mm 0mm},clip]{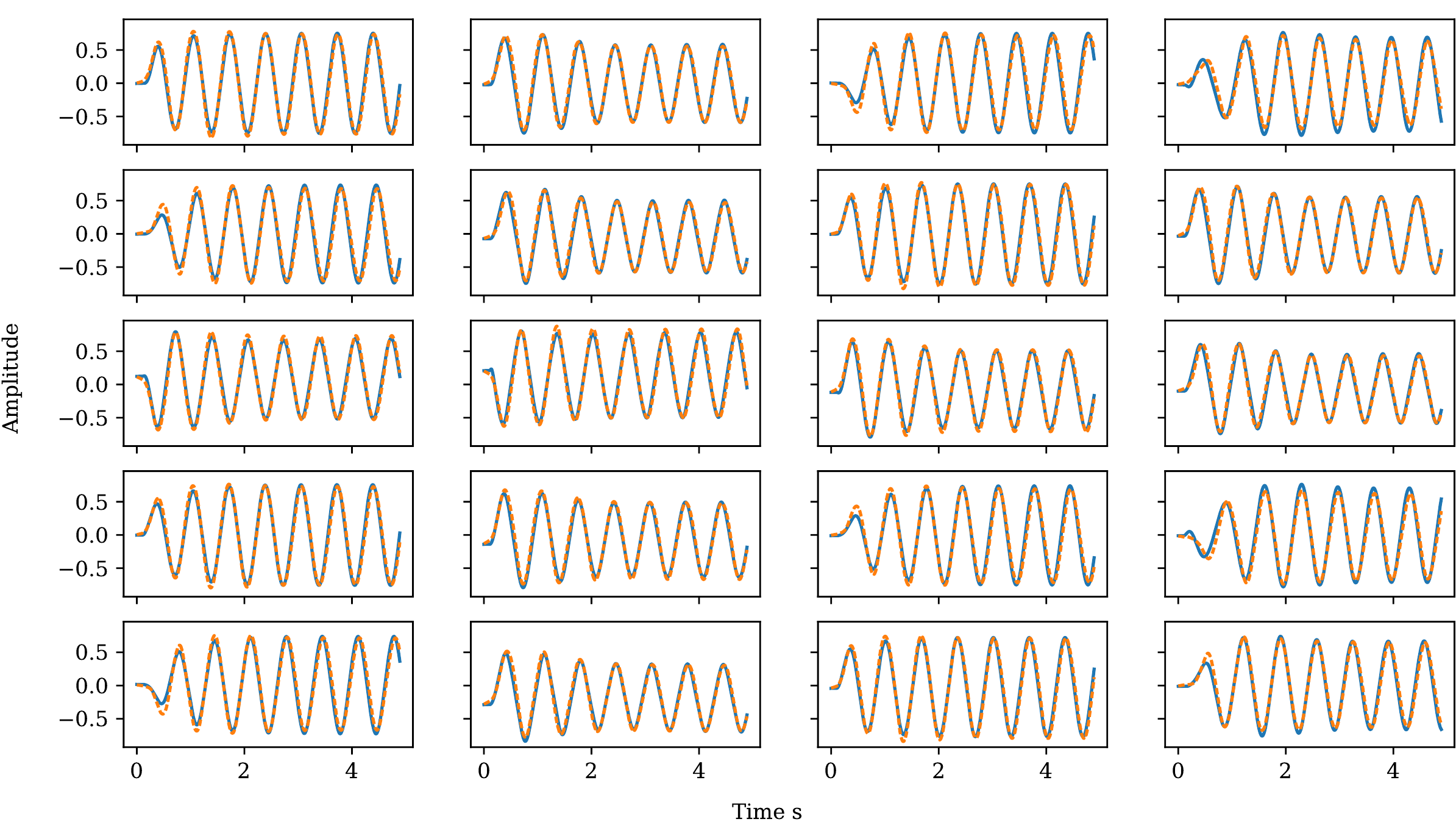}}
    \caption{Comparison of the true response (blue line) to the response predicted by the ROM (orange dashed line) at all DOF for a new forcing time history for the 108 DOF Bouc-Wen frame system.}
    \label{fig:ROM_BW}
\end{figure}

\section{Physical Interpretation of NNMs}\label{sec:Retained NNMs}
In order to visualise the dynamics captured by the retained NNMs in the autoencoder, the time-series of different DOFs were plotted as reconstructed using autoencoders with increasing numbers of bottleneck layer nodes, and hence, number of NNMs. A study was made making use of the 20 DOF single nonlinearity system wherein autoencoders were trained with an increasing number of bottleneck units and the reconstructed time-series for each compared to the true value. Figures 28 and 29 show the response as reconstructed from an increasing number of retained NNMs for the 1st and 10th DOF respectively for the first 10 seconds of the time-series. As expected, in both cases an increasing number of retained NNMs results in a higher fidelity for the reconstruction of the responses. It is also evident that an increasing number of NNMs appears to result in an increasing amount of high frequency information being retained. In this case, both the 1st and 10th degrees of freedom are poorly represented with few NNMs and both achieve high fidelity with more than 7 NNMs.

Figures 30 and 31 show the response reconstructed from an increasing number of NNMs at the 1st and 10th DOF respectively, focused between the 40th and 50th second of the analysis. For this part of the time-series, the transient behaviour of the system is less prevalent. As such, the response at the 10th DOF achieves a much better fidelity with relatively few retained NNMs in comparison to the 1st DOF. This shows that the response at the 10th DOF is more dominated by lower frequency response whilst at the 1st DOF higher frequency response is more important.

\begin{figure}[h!]
    \centering
    \includegraphics[width=120mm]{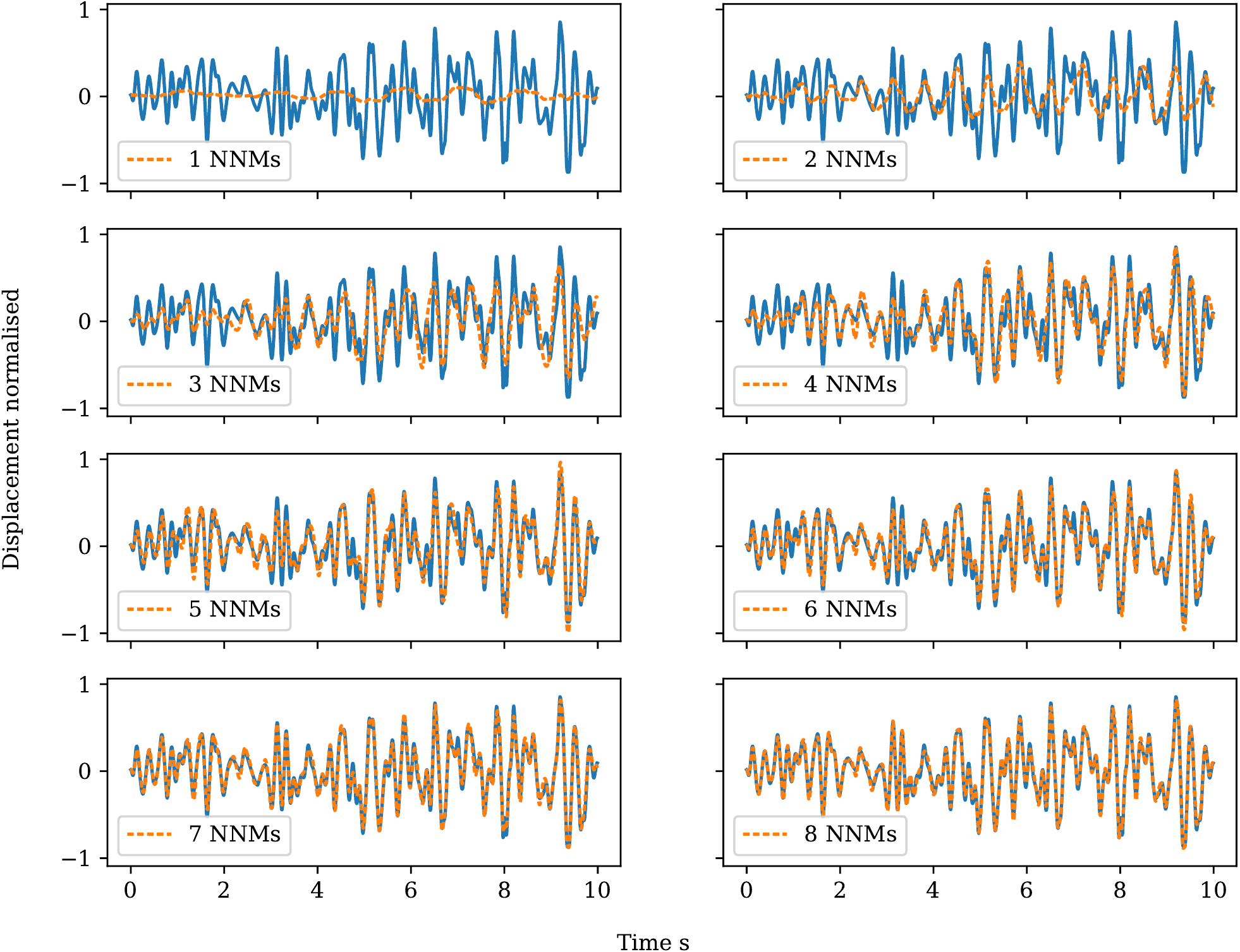}
    \caption{Response at 1st DOF between 0-10 seconds as recreated from increasing number of NNMs.}
    \label{fig:1stDOF1_10}
\end{figure}

\begin{figure}[h!]
    \centering
    \includegraphics[width=120mm]{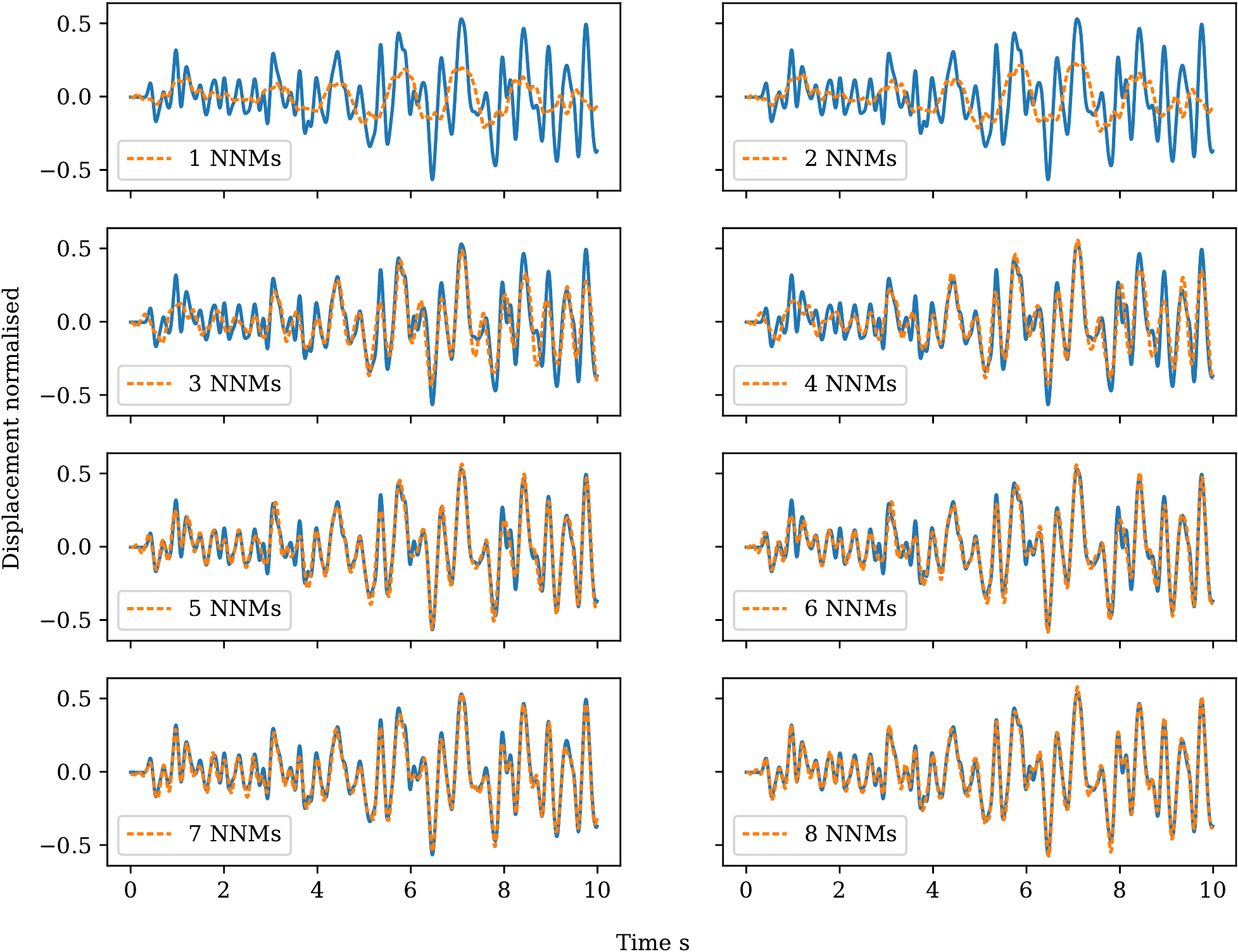}
    \caption{Response at 10th DOF between 0-10 seconds recreated from increasing number of NNMs.}
    \label{fig:10thDOF1_10}
\end{figure}

\begin{figure}[h!]
    \centering
    \includegraphics[width=120mm]{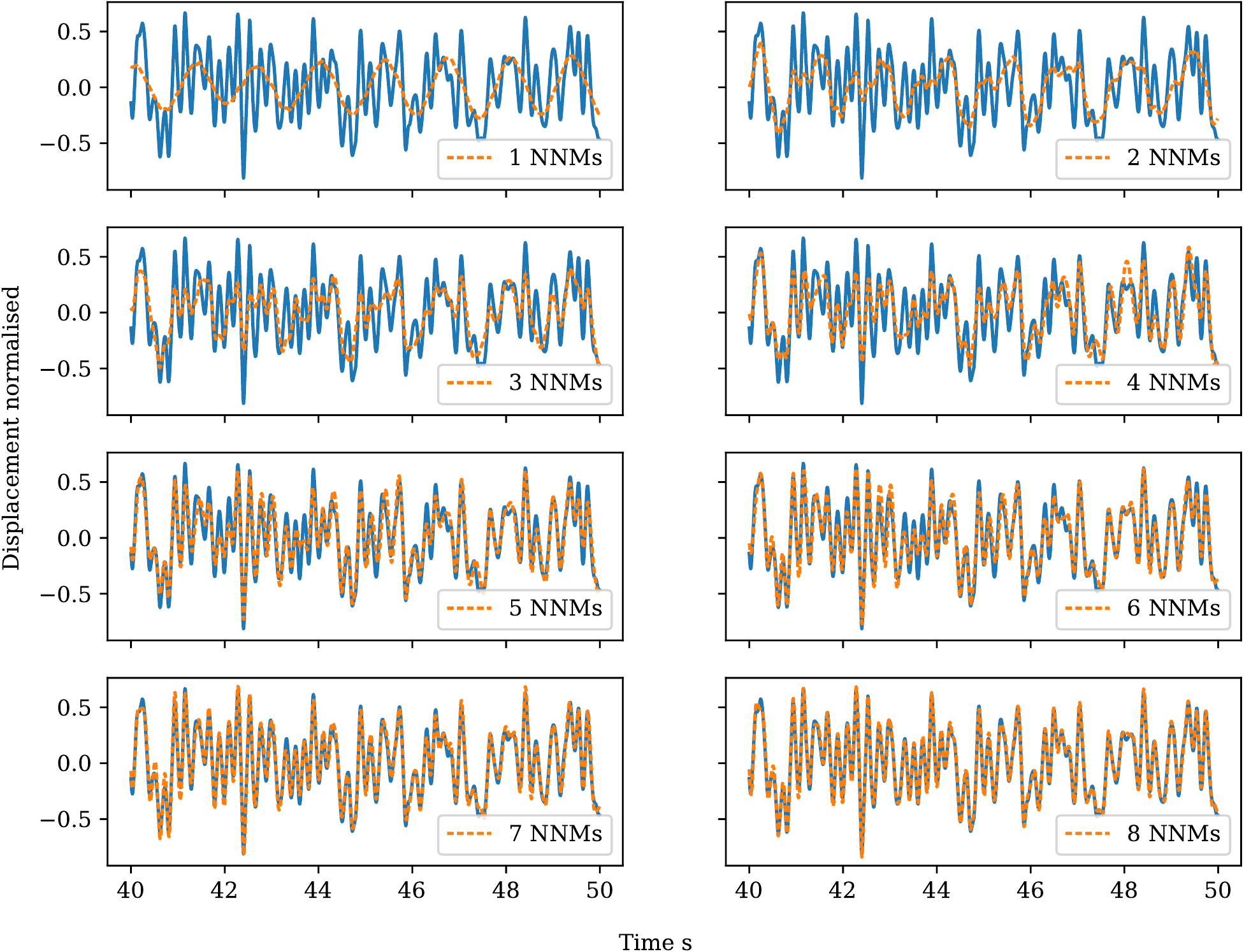}
    \caption{Response at 1st DOF between 40-50 seconds as recreated from increasing number of NNMs.}
    \label{fig:1stDOF_40_50}
\end{figure}

\begin{figure}[h!]
    \centering
    \includegraphics[width=120mm]{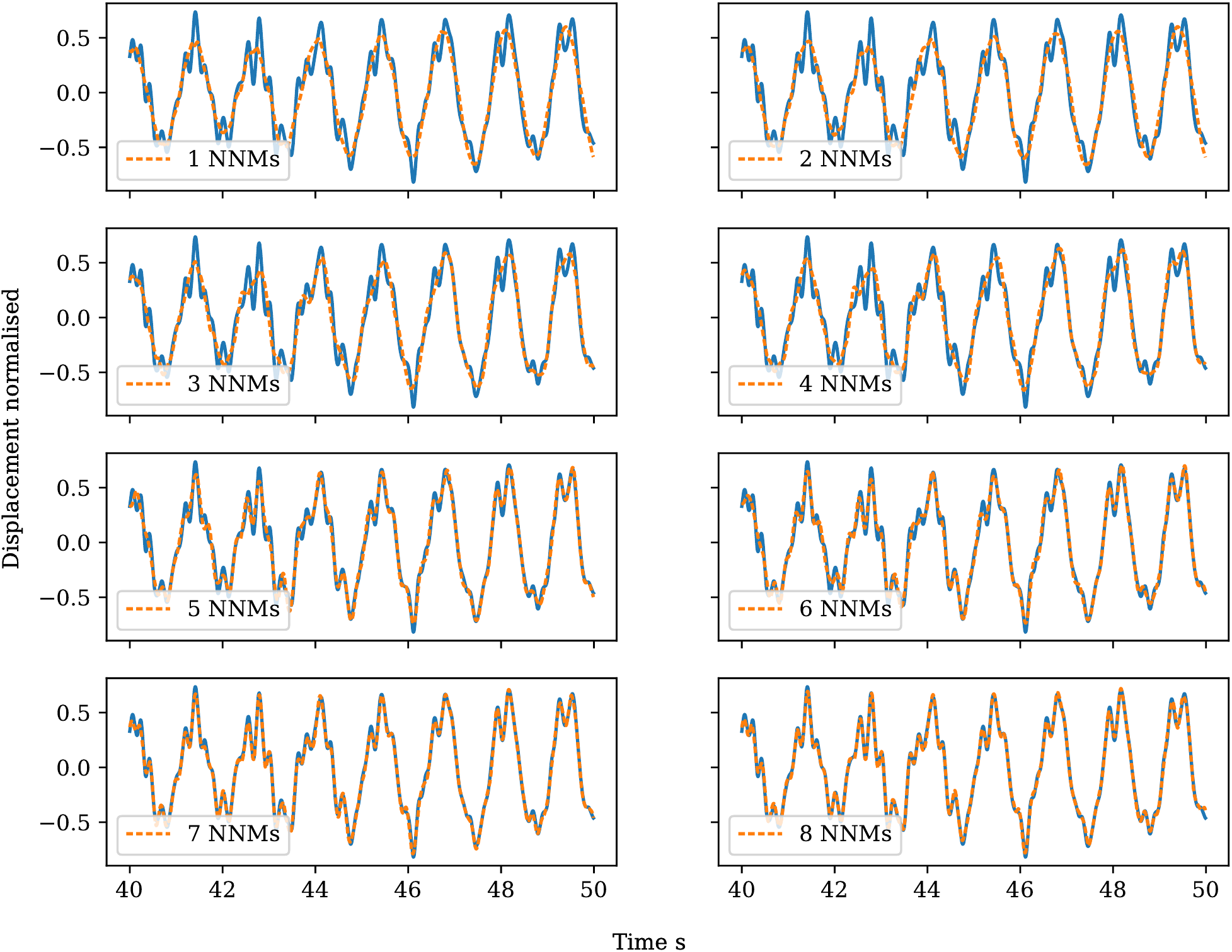}
    \caption{Response at 10th DOF between 40-50 seconds as recreated from increasing number of NNMs.}
    \label{fig:10thDOF_40_50}
\end{figure}

\clearpage

\section{Conclusions and Further Work}\label{sec:conclusions}

This work presents a novel reduced order modelling method relying on availability of input and output data for constructing a surrogate able to reproduce the response of a nonlinear structural systems under ``unseen" forcing time histories. An Autoencoder is employed for deriving a latent space, attempting to recover NNM-like quantities that are exploited as a reduction basis. The extraction of these NNMs from simulated data using autoencoders is discussed, along with the creation of a predictive model using LSTM neural networks. The modelling procedure and its efficacy are demonstrated on various nonlinear systems of differing complexity and size.

The demonstrated method features numerous advantages:
\begin{itemize}
\item No a priori assumptions are posed on the type of underlying nonlinearity, or the required magnitude, rendering the approach applicable to general-type nonlinear systems.
\item Secondly, due to the nonlinear nature of the NNMs, they should, if sufficient training data is used, be generally valid over the whole state trajectory of a system, as opposed to methods that rely on linear projections. 
\item It is further demonstrated via the employed examples that the method is applicable across diverse types of structural nonlinearity. 
\end{itemize}

It is worthy of note, however, that as a data-based method, this technique is naturally limited to the modelling of only those dynamics, which have been excited by the forcing cases considered in training. This has the obvious drawback that dynamics, which are not excited in the training dataset, will naturally not be captured by the ROM. However, as long as the forcing applied is representative, the method will capture only those dynamics which are relevant to the system without including any unnecessary complexity.

As with any reduced order modelling method, the time/computation savings provided by the method are of equal importance to accuracy. The example systems considered herein, however, are only intended to serve as a proof of concept for the proposed method. The 20 DOF system, in particular, serves as an educational example of rather low dimensionality in comparison to complex nonlinear FE-based models. This implies, that the full order model of the 20 DOF example already requires rather low computational toll, which does not allow for the benefits of reduction in terms of a pure comparison of computational time to be observed. In the case of the 20 DOF system with a single nonlinearity, the required time to simulate 10 seconds of response with the full order model was $1.10 s$, whilst the ROM required $1.54 s$. In the case of the 20 DOF system with multiple nonlinearities, the required time to simulate 10 seconds of response with the full order model was $1.10 s$, whilst the ROM required $6.8 s$. There may well be considerable potential to reduce the size of the LSTM network used until an advantage can be observed in these smaller systems. However, as with the majority of ROM methods, the benefits are more evident when considering larger systems. The Bouc-Wen frame structure considered was of 108 DOFs. Whilst this is still a relatively limited size example compared with full scale finite element models, the benefits of the ROM procedure in terms of acceleration become this time apparent. In the case of the Bouc-Wen frame model the computational time to simulate 5 seconds of response was $0.82 s$ for the ROM as oppose to $7.3 s$ for the full order model.

Interesting insights were also drawn with regards to the physical interpretation of the NNMs extracted by the autoencoders. It was demonstrated that an increase in the number of NNMs tends to conform to an increased amount of high frequency information preserved by the autoencoder. It is worth considering how this ROM method can be extended to high dimensional systems. This issue is different when considering high dimensional output i.e., many DOFs at which the response is recorded, and high dimensional input, i.e., forces applied at several DOFs. When considering systems with a high dimensional output, this affects the autoencoder and the extraction of NNMs. Normally, this high dimensional data would result in an autoencoder with a very large number of parameters. This can result in issues with regards to both training time and the amount of training data required for such a network. The use of convolutional neural networks (CNNs) could largely alleviate these issues. CNNs make use of layers which perform a convolutional operation of the inputs with a set number of fixed size filters. Only the parameters of these filters must be learned. This effectively decouples the number of parameters to be trained from the dimensionality of the data. CNNs are effectively and widely used in high dimensional image recognition tasks \cite{RawatWang} and also used in the context of model reduction by Lee and Carlberg \cite{LEE2020108973}. When considering high dimensional inputs, however, the effect largely falls on the regression model constructed to model the dynamics of the system, in that every additional forcing is an additional exogenous input to be considered. This could result in the regression problem becoming much more difficult if a very large number of inputs were used. There could, however, under such circumstances be scope to perform some kind of reduction on the input space.

Future developments will consider the expansion of the method to large scale FE systems and a full investigation of the computational benefits possible with such a method. In transferring to larger scale systems, it is expected that methods such as Convolutional Autoencoders will become useful in order to prevent the number of parameters exploding with increased number of degrees of freedom. A detail worthy of further consideration lies in the contribution of statistical independence of the extracted variables in justifying their classification as NNMs \cite{Worden2016}. It could be of interest to investigate specific forms of the autoencoder that can enhance this effect. This could be examined via use of the variational autoencoder, wherein the latent variable is encouraged to take the form of a diagonal Gaussian distribution \cite{Kingma2014}, or investigation of the recent work of \cite{ladjal2019pcalike}, wherein the correlation of the latent variables is also minimised during network training. This is left for future work.

\section*{Data Availability Statement}
 All data, models, or code that support the findings of this study are available from the corresponding author upon reasonable request.

\section*{Acknowledgements}
This work was carried out as part of the ITN project DyVirt and has received funding from the European Union’s Horizon 2020 research and innovation programme under the Marie Skłodowska-Curie grant agreement No 764547

%
%

\newpage
\bibliography{bib}

\end{document}